\documentclass[aps,pra,showpacs,nofootinbib,twocolumn]{revtex4-1} 

\usepackage[latin1]{inputenc}
\usepackage{amsthm}
\usepackage{amssymb}
\usepackage{amsmath}
\usepackage{bbold}
\usepackage[pdftex]{hyperref}
\usepackage{graphicx}
\usepackage{dsfont}
\usepackage{xcolor}
\usepackage{float} 

\newcommand{\bra}[1]{\left\langle #1 \right|}
\newcommand{\ket}[1]{\left| #1 \right\rangle}

\newcommand{\ketbra}[1]{\left|#1\middle\rangle\middle\langle#1\right|}
\newcommand{\abs}[1]{\left|#1\right|}
\newcommand{\ie}{i.e.,}
\newcommand{\ba}{\begin{eqnarray}}
\newcommand{\ea}{\end{eqnarray}}
\newcommand{\Tr}{\text{Tr}}
\newcommand{\norm}[1]{\left\lVert#1\right\rVert}

\renewcommand{\leq}{\leqslant}
\renewcommand{\geq}{\geqslant}

\begin{document}

\title{Algorithmic construction of local models for entangled quantum states: optimization~for~two-qubit~states}

\author{Mathieu Fillettaz}
\thanks{These authors contributed equally to this work.}
\author{Flavien Hirsch}
\thanks{These authors contributed equally to this work.}
\author{S\'ebastien Designolle}
\author{Nicolas Brunner}
\affiliation{D\'epartement de Physique Appliqu\'ee, Universit\'e de Gen\`eve, 1211 Gen\`eve, Switzerland}

\date{\today}  

\begin{abstract}
  The correlations of certain entangled quantum states can be fully reproduced via a local model. We discuss in detail the practical implementation of an algorithm for constructing local models for entangled states, recently introduced by Hirsch et al.~[Phys.~Rev.~Lett.~{\bf 117}, 190402 (2016)] and Cavalcanti et al.~[Phys.~Rev.~Lett.~{\bf 117}, 190401 (2016)]. The method allows one to construct both local hidden state (LHS) and local hidden variable (LHV) models, and can be applied to arbitrary entangled states in principle. Here, we develop a systematic implementation of the algorithm, discussing the choice of the free parameters. For the case of two-qubit states, we design a ready-to-use procedure. This allows us to construct LHS models (for projective measurements) that are almost optimal, as we show for Bell diagonal states, for which the optimal model has recently been derived. Finally, we show how to construct fully analytical local models, based on the output of the convex optimization procedure.
\end{abstract}

\maketitle

\section{Introduction}

Quantum experiments consisting of local measurements performed by distant parties on a shared entangled quantum state can produce nonlocal correlations, i.e., probability distributions which admit no local explanation \cite{Bell64,review}. These distributions can be witnessed by the violation of a Bell inequality, and are useful resources, for example in device-independent randomness expansion and cryptography \cite{ABG07,Col,PAM10,Rotem}. Such distributions can be produced by local measurements on entangled states, the presence of entanglement being in fact a necessary condition for nonlocality within quantum mechanics.

However, entanglement and nonlocality are inequivalent phenomena, as there exist entangled quantum states which cannot give rise to nonlocal distributions. This was first shown by Werner \cite{Werner89}, who presented a class of entangled states whose statistics can be reproduced by a local hidden-variable (LHV) model, considering arbitrary projective measurements. This was later extended to more general positive-operator valued measures (POVMs) \cite{Barrett02}, to other classes of states \cite{Almeida07,Acin06,GHNL,Augusiak_review}, and also to multipartite states \cite{Toth06,Augusiak15,Bowles16b}.

Another manifestation of the non-classical character of entanglement is that of EPR steering \cite{Wiseman07,paul_review}. This type of correlations, strictly weaker than nonlocal ones, captures the fact that Alice can remotely steer Bob's state by measuring half of an entangled state. Equivalently, steerable correlations are those that cannot be explained with a local hidden-state (LHS) model, a particular class of LHV models. Steering turns out to be intermediate, and strictly inequivalent, to both entanglement and nonlocality \cite{Quintino15}. Classes of entangled states that are unsteerable, i.e., admitting a LHS model were presented, see e.g.~\cite{Wiseman07,paul_review,Bowles14,Jevtic15,Quintino15,Bowles15,Bowles16,Nguyen16,Miller18}. Moreover, steering is intimately connected to the notion of joint measurability \cite{Uola14,Quintino14,Uola15} which provides further applications of LHS models.

More generally, it is not understood which entangled states can give rise to nonlocality or steering. Beyond its fundamental character, this question is also natural in the context of Bell experiments and nonlocality-based applications. One of the main hurdles to this problem is the fact that constructing local models (LHS or LHV) is challenging in general. Indeed, all the above mentioned examples of entangled states admitting a local model feature a high degree of symmetry, which greatly simplifies the construction of the model.

Recently, a general method was developed in order to address this question \cite{Hirsch16,Cavalcanti16}. Importantly, this method is algorithmic and does in principle allow one to construct a local model (LHS or LHV) for an arbitrary target state (given that the state admits such a model). The method can be implemented as a sequence of tests, with growing computational complexity. Importantly this is of practical interest, as even low complexity tests give interesting results. The method already found diverse applications \cite{Sainz15,Hirsch16b,Nagy16,Bavaresco17,Hirsch17,Toth18,Orieux17}, for instance, for demonstrating that measurement incompatibility does not lead to Bell nonlocality in general \cite{Bene18,Hirsch18}.

In the present work, we discuss in detail the implementation of this algorithmic method for constructing LHS and LHV models. While Refs \cite{Hirsch16,Cavalcanti16} developed the general idea of the algorithm, they did not provide an explicit implementation; all illustrative examples were essentially treated case by case. Here, we present a systematic implementation of the algorithm, in a ready-to-use form. We focus mainly on the simplest case of two-qubit entangled states with projective measurements, for which case we construct an efficient and versatile procedure. In particular, for LHS models, we construct models that are very close to being optimal. To do so, we benchmark our method for Bell diagonal states, as their steering properties were recently fully characterized \cite{Jevtic15,Nguyen16,Zhang17,Yu17}.

The article is organized as follows. We first introduce the concepts and notations in Section \ref{prel}. Next, we describe the algorithm in Section \ref{recall}. In Section \ref{opt} we discuss how to improve the efficiency of the algorithm, characterizing the role of each free parameter. While the technique is inherently numerical, we show in Section \ref{analytical} how to make all results fully analytical. We study the performance of our algorithm for several classes of two-qubit entangled states in Section \ref{results}. Finally, we conclude in Section \ref{conclusion}.

\section{Preliminaries} \label{prel}

Consider two distant observers, Alice and Bob, sharing an entangled state $\rho$. Alice has access to a set of measurements $\{A_{a|x} \}$ ($A_{a|x}\geq 0$ and $\sum_a A_{a|x} = \mathds{1}$ for all $x$), and Bob has $\{B_{b|y}\}$ (with similar conditions). Here, $x$ and $y$ denote the measurement choice, and $a$ and $b$ the outcomes. The resulting statistics is given by
\begin{align} \label{pQ}
  p(ab|xy) = \Tr ( A_{a|x} \otimes B_{b|y} \; \rho ).
\end{align}
Such statistics is said to be nonlocal when
\begin{align} \label{NL}
  p(ab|xy) \neq \int \pi(\lambda) \; p_A (a|x,\lambda ) \; p_B (b|y,\lambda) \; d\lambda
\end{align}
for any variable $\lambda$, distributed with density $\pi(\lambda)$, and for any local response distributions $p_A (a|x,\lambda )$ and $p_B (b|y,\lambda)$. That is, the statistics cannot be reproduced using a LHV model. In this case, the state $\rho$ is said to be nonlocal, as witnessed by the fact that $p(ab|xy)$ violates (at least) one Bell inequality \cite{review}.

Oppositely, we say that a state $\rho$ is local if its statistics admits a LHV model. That is, if one can find a shared variable and local distributions such that
\begin{align} \label{LHV}
  \Tr ( A_{a} \otimes B_{b} \; \rho ) = \int \pi(\lambda) \; p_A (a|\{ A_{a} \},\lambda ) \; p_B (b|\{ B_{b} \},\lambda) \; d\lambda
\end{align}
for all measurements $\{A_{a}\}$ and $\{B_{b}\}$. Here, one can consider different sets of measurements. For any set of measurements that is finite, methods based on linear and semi-definite programming can be used, see e.g.~\cite{review,Pusey13,Paul14,paul_review,Terhal}. The main challenge however consists in constructing local models considering sets of measurements that are continuous, as for instance, the set of all projective measurements, or the set of all POVMs. Such a model was first constructed by Werner \cite{Werner89} for a specific class of entangled states that he introduced. While Werner focused on projective measurements, this was later extended by Barrett to general POVMs. For a review, see Ref.~\cite{Augusiak_review}.

A specific class of local models are LHS models, of the form
\begin{align}
  \Tr ( A_{a} \otimes B_{b} \, \rho )  = \int \pi(\lambda) \; p_A (a|\{ A_{a} \},\lambda ) \; \Tr ( B_{b} \sigma_\lambda ) \; d\lambda  \,.
\end{align}
The specificity of these models is that the hidden variable on Bob's side is a quantum state $\sigma_\lambda$, while Alice has a fully classical description of this state. Note that both Werner's and Barrett's models are in fact of this form which shows the existence of entangled states that admit a LHS model for general POVMs \cite{Quintino15}.

Entangled states for which such a model does not exist are steerable. One can then find a set of measurements for Alice $\{A_{a|x} \}$ that leads to steering. Note that in a steering test one assumes that Bob's measurement device is well-characterized. For simplicity, one can take Bob's measurements to be tomographically complete, which allows him to reconstruct the assemblage
\begin{align}
  \sigma_{a|x} = \Tr_A (A_{a|x} \otimes \mathds{1} \; \rho),
\end{align}
i.e., the collection of conditional states, remotely prepared by Alice's measurements. Steering is detected whenever the assemblage cannot be decomposed as
\begin{align}
  \sigma_{a|x}  \neq \int \pi(\lambda) \; p_A (a|x,\lambda ) \sigma_\lambda  \; d\lambda \,,
\end{align}
considering any possible shared variable $\lambda$ and local distribution $p_A (a|x,\lambda )$. In this case, the assemblage will lead to violation of (at least) one steering inequality.

\section{The algorithmic method} \label{recall}

We start by discussing the algorithmic method presented in \cite{Hirsch16,Cavalcanti16}, and reviewing the protocols for constructing local models. The method is in general applicable to any entangled state, and will eventually return the local model (given that such a model exists). While this allows one to construct both LHS and LHV models in general, we will focus on the former in the main text, for clarity; details on LHV models will be given in Appendix \ref{app:lhv}.

The main idea behind the method is to map the initial problem to a much simpler one. The initial problem is challenging as it considers sets of measurements that are continuous. The final problem will turn out to be much simpler as it considers only finite sets of measurements, in which case standard methods can be applied efficiently. A continuous set of measurements that are slightly noisy can be entirely captured by considering only a finite set of noiseless measurements. 

Consider an initial target entangled state $\rho$, and a set of measurements $\{A_{a} \}$. One can then construct a different state $\chi$ whose statistics for a set of noisy measurements $\{A_{a}^{\eta}\}$ is exactly equivalent to that of performing noiseless measurements $\{A_{a} \}$ on $\rho$. Given that $\chi$ admits a local model for noisy measurements $\{A_{a}^{\eta}\}$ (which can be checked by considering only a finite set of noiseless measurements), we obtain that $\rho$ admits a local model for measurements $\{A_{a}\}$. For an illustrative example, we refer the reader to Ref.~\cite{Hirsch16} page 2.

More formally, we first define the following map, which allows us to define the noisy measurements from the initial ones $\{A_{a}\}$:
\begin{equation}\label{map}
  \Phi^{\eta}(A_{a})  = \eta A_{a} + (1-\eta) \Tr(\xi A_a) \mathds{1} \equiv A_{a}^{\eta}
\end{equation}
where $0 \leq \eta \leq 1$ and $\xi$ is a density matrix. Note that $\Phi^{\eta}$ is unital, and thus maps POVMs into valid POVMs. Next, observe that the statistics of these noisy measurements on a given state $\chi$ are equivalent to the statistics of a noisy state $\chi^{\eta}$ and noiseless measurement, i.e.

\begin{equation} \label{equiv}
  \Tr( A_a^\eta \otimes B_b\, \chi) = \Tr(A_a \otimes B_b\, \chi^\eta).
\end{equation}
Note that $\chi^\eta$ is found by applying the dual map $\Phi_{*}^\eta$ on Alice's side, namely,

\begin{equation}\label{dualmap1}
  \chi^\eta = \Phi_{*}^{\eta}(\chi)  = \eta \chi +  (1-\eta)  \xi \otimes \chi_B
\end{equation}
where $\xi$ is the density matrix defining the map, see Eq.~\eqref{map}, and $\chi_B = \Tr_A(\chi)$ is the reduced state of Bob.

The final step consists in proving that the left-hand side of Eq.~\eqref{equiv} admits a LHV model, which implies that the right-hand side also does. This can be done by considering only a finite set of measurements $\{ M_{a|x} \}$, given that any noisy measurement $\{A_a^\eta\}$ can be expressed as a convex combination of elements of $\{ M_{a|x} \}$. This can be understood geometrically. The $\{ M_{a|x} \}$ forms a polytope (in the space of measurements). If this polytope fully contains the entire set of noisy measurements $\{A_a^\eta\}$, then any of the latter can be decomposed as a convex mixture of elements of $\{ M_{a|x} \}$. Next, if we can ensure that the statistics resulting from the finite set of measurements $\{ M_{a|x} \}$ on $\chi$ admit a LHS model (which can be done e.g.~via semi-definite programming \cite{paul_review}), it follows by linearity that the same holds for all noisy measurements $\{A_a^\eta\}$. Finally, taking $\rho= \chi^\eta$ we obtain a LHS model for $\rho$ for all measurements $\{A_{a}\}$.

In practice, the algorithm can be implemented in the following way. Given a target state $\rho$, we define the class of states
\ba \label{rho_q}
\rho_q = q \rho + (1-q) \rho_{sep}
\ea
with $0\leq q \leq 1$, and where $\rho_{sep}$ is a separable state (hence unsteerable). We will aim at finding the maximum value of $q$, the maximal visibility $q^*$, such that $\rho_{q^*}$ admits a LHS model. Choose a finite set of measurements $\{ M_{a|x} \}$ such that all noisy measurements $\{A_{a}^{\eta}\}$ can be written decomposed as convex combinations of $\{ M_{a|x} \}$; equivalently, this fixes the ``shrinking factor'' $\eta$. Then, run the following semi-definite program:
\medbreak

{\bf LHS Protocol}
\ba \label{LHSprot}  \text{find  } & & q^* = \max  q  \\
\text{s.t.  }   & & \Tr_A(A _{a|x} \otimes \mathds{1} \, \chi) = \sum_\lambda \sigma_\lambda D_\lambda(a|x)  \quad  \forall a,x,   \nonumber  \\ \nonumber
& &   \eta \chi + (1-\eta) \xi \otimes \chi_B = \rho_q, \quad \sigma_\lambda \geq 0 \quad  \forall \lambda.
\ea
The SDP optimization variables are (i) the positive matrices $\sigma_\lambda$ and (ii) a hermitian matrix $\chi$.\footnote{Note that $\chi$ does not need to be positive in general; see Section \ref{rank3} for a practical example where taking $\chi$ non-positive is useful.} This SDP must be performed considering all possible deterministic strategies for Alice $D_\lambda(a|x)$, of which there are $n= k^m$, where $m$ denotes the number of measurements of Alice and $k$ the number of outcomes. Hence $\lambda = 1,\ldots,n$. If the optimization returns a maximum of $q^*=1$, then $\rho$ admits a LHS model. If $q^*<1$, then we have at least shown that $\rho_{q^*}$ admits a LHS model.

More generally, we can define a sequence of tests. Start from a finite set of measurements $\{ M^1_{a|x} \}$, with shrinking factor $\eta_1$. This is the initial setting ($j=1$) of the following iterative process:

\medbreak

Step 1: Take measurements $\{ M^j_{a|x} \}$ and run the LHS protocol \eqref{LHSprot}.
\begin{itemize} \label{iterative}
  \item If $q^* \geq 1$ the algorithm stops. The state $\rho$ then admits a LHS model, which can be reconstructed explicitly from the values of the SDP variables.
  \item If $q^* < 1$, we construct a new finite set of measurements $\{ A^{j+1}_{a|x} \}$ with associated shrinking factor $\eta_{j+1} > \eta_j$. Below we will discuss how to construct such a new set starting from the previous one $\{ A^j_{a|x} \}$ and adding measurements.
\end{itemize}
Step 2: Set $j=j+1$ and go back to step $1$.

\medbreak

In the limit $k \rightarrow \infty$, this algorithm converges, in the sense that it will return $q^*=1$ if $\rho$ admits a LHS model. 

The same ideas lead an algorithm for constructing LHV models \cite{Hirsch16,Cavalcanti16}. Similarly to the above presentation, we discuss the case of LHV models in Appendix \ref{app:lhv}.

\section{Optimization of the algorithm with focus on qubits} \label{opt}

In the previous section, we reviewed the algorithm for constructing LHS models. It appears clearly that there are a number of parameters in the method to be set initially by the user. In general, we observe that the performance of the algorithm is considerably improved by a judicious choice of these parameters. Moreover, there are additional parameters that can be introduced in order to further boost the performance.

The goal of this section is to give insight as to how the algorithm can be optimized in practice. Specifically, we discuss the following points:
\begin{enumerate}
  \item Choice of the noise map $\Phi^{\eta}$ in Eq.~\eqref{map}, i.e., definition of the density matrix $\xi$
  \item Choice of the finite set of measurements $\{ M_{a|x} \}$ and computation of the shrinking factor $\eta$
  \item Adding auxiliary states
  \item Removing redundant constraints in the SDP
  \item Selecting deterministic strategies
\end{enumerate}
We will discuss each of these points. While we focus on the case of LHS models for two-qubit states, we believe that these ideas will also improve performance in more general cases.

\subsection{Choice of the noise map}

The first parameter to set is the noise map $\Phi^{\eta}$ defined in Eq.~\eqref{map}. Specifically, we need to choose the density matrix $\xi$, which defines the map. In general, we observe that the best choice of $\xi$ depends on the input state. While the simplest, and probably most natural choice, namely to set $\xi = \mathds{1}/2$ (here for qubits), gives relatively good results in most cases, it is in general not optimal.

Based on many examples, we conjecture that the optimal choice of $\xi$ is the following.
Given a target state of the form as given in Eq.~\eqref{rho_q} setting $\xi = \Tr_{B}(\rho_{sep})$ appears to be the best choice. 

\subsection{Choice of the finite set of measurements}

The choice of the finite set of measurements used for approximating the entire (continuous) set of measurements is very important. This can be understood intuitively for qubit projective measurements. Here, a measurement is characterized by a Bloch vector. Taking a finite set with few measurements gives only a rough approximation of the entire Bloch sphere; hence the shrinking factor will be relatively small. On the other hand, taking a large number of measurements, well distributed over the sphere, provides a good approximation of the sphere, hence a shrinking factor close to one. Since obviously sets with more and more measurements become much more difficult to handle computationally, it is important to find the right balance.

Another important point is the computation of the shrinking factor. Note that in general, given a choice of measurements, the shrinking factor $\eta$ will still depend on the choice of the noise map. Below we give two methods to compute efficiently the shrinking factor for finite sets of qubit projective measurements: (i) when $\xi = \mathds{1}/2$ , (ii) for arbitrary $\xi$. We also briefly discuss the case of general qubit POVMs. Finally, note that Ref.~\cite{Hirsch16} provided a general method for computing the shrinking factor, yet the methods presented here are more efficient for the case of qubits.

\subsubsection{Isotropic map\texorpdfstring{, $\xi=  \mathds{1}/2$}{}} \label{projqubit}

Applying the resulting map \eqref{map} to the entire set of projective measurements leads to the following (continuous) set of noisy qubit measurements:
\ba
\{ A^{\eta} | A^{\eta} = \eta A + (1-\eta) I \}
\ea
where $A$ is a Pauli observable and $I = \{ \mathds{1} / 2,\mathds{1} / 2 \}$. We can write
\ba
A = \{ A_{+}, A_{-} \} , \; A_{\pm} = \frac{ \mathds{1} \pm \hat{v} \cdot \vec{\sigma} }{2}
\ea
where $\vec{\sigma} = \{ \sigma_x, \sigma_y, \sigma_z \}$ contains the Pauli matrices and $\hat{v}$ is a normalized Bloch vector. For any $A^{\eta}$, we therefore have
\ba
A^{\eta}= \{ A^{\eta}_{+}, A^{\eta}_{-} \} , \; A^{\eta}_{\pm} = \frac{ \mathds{1} \pm \eta \hat{v} \cdot \vec{\sigma} }{2} .
\ea
This set represents a ``shrunk Bloch sphere'' of radius $\eta$. Thus, given a finite set of projectors $\{ M_x \}$ (with Bloch vectors $\{ \hat{v}_x \}$), the shrinking factor is simply the radius of the largest sphere that fits inside the polyhedron generated by $\{ \hat{v}_x \}$. This radius can be computed with arbitrary precision for any polyhedron by characterizing its facets, the radius of the inscribed sphere being then the distance from the center of the sphere to the closest facet. Since the facet enumeration problem is very efficient in dimension three, using polyhedrons with many vertices (more than a thousand) is feasible. Note also that several families of polyhedra over the sphere are known, in which case the shrinking factor is obtained analytically.

\subsubsection{General map} \label{eta_m}

We now consider a general noise map $\Phi^{\eta}$ (see Eq.~\eqref{map}), i.e., $\xi= \left(\mathds{1} + \vec{u}\cdot\vec{\sigma}\right)/2$ is now an arbitrary qubit state. Applying the map to all projective measurements, we obtain a set of noisy binary measurements, with POVM elements
\begin{equation}
  A^{\eta}_\pm = \left(\frac{1}{2} \pm (1{-}\eta)\frac{\vec{u}\cdot\hat{v}}{2}\right)\mathds{1} \pm \frac{\eta\hat{v}\cdot\vec{\sigma}}{2} \,.
\end{equation}
Since $A^{\eta}_+ + A^{\eta}_- = \mathds{1}$, we can focus on the first POVM element. The POVM is then characterized by the vector

\begin{equation}\label{eq:vectors_4d}
  \textbf{v}^{\eta}_+ = \left\lbrace\left(\frac{1}{2} + (1{-}\eta)\frac{\vec{u}\cdot\hat{v}}{2}\right),\frac{\eta \hat{v}}{2}\right\rbrace
\end{equation}
in the four-dimensional space spanned by $\left\lbrace\mathds{1},\vec{\sigma}\right\rbrace$.

Notice that when $\xi = \mathds{1}/2$, the first component of this vector is always equal to $1/2$ and can thus simply be ignored. The problem is then reduced to a three-dimensional problem, hence the Bloch representation is sufficient.

In the general case, the problem is now the following. Given a finite set of measurements $\{ M_{a|x} \}$, we can represent each element of the set by a vector in $\mathds{R}^{4}$. We obtain a polytope, of which we can find the facets. Each facet is characterized by a vector $\textbf{F}_{j} \in \mathds{R}^{4}$ and a real number $b_j$. A vector $\textbf{p} ~\in~ \mathds{R}^{4}$ is inside the polytope if and only if

\begin{equation}\label{eq:cond_shrink}
  \left(\textbf{F}_{j},\textbf{p}\right) \leq b_{j} \qquad \forall j = 1,\ldots,N_{F},
\end{equation}
where $N_F$ denotes the number of facets.


Our task now is to determine the largest value of $\eta$ such that
\begin{equation}
  \left(\textbf{F}_j,\textbf{v}^{\eta}_+\right) \leq b_j 
\end{equation}
for all noisy POVMs $\textbf{v}^{\eta}_+$ and all facets ($j = 1,\ldots,N_{F}$). For each facet $j$, one can actually find analytically the maximal value of $\eta$, $\eta^*_j$, by solving a quadratic equation; see Appendix \ref{app:shr2}. Finally, the shrinking factor $\eta^*$ is obtained by taking the minimum over all values $\eta^*_j$.

\subsubsection{General qubit POVMs}

Here, we can restrict to the case of four-outcome POVMs, as any qubit POVM can be viewed as classical post-processing of some four-outcome POVMs \cite{Dariano05}. Each POVM element can be expressed in the Pauli basis $ A_a= v_a \mathds{1} +
\textbf{v}_a \cdot \vec{\sigma}$, 
hence represented by a four-dimensional vector. Thus the full POVM is characterized by a vector in
$\mathds{R}^{12}$, taking normalization into account. This makes the problem much more difficult compared to the case of binary measurements. First, running facet enumeration algorithms is here much more costly.
Also, finding the shrinking factor for each facet can no longer be solved analytically, but can be treated as an SDP. In practice, this problem can still be solved for certain cases, and was used successfully in certain applications; see \cite{Hirsch16b,Hirsch18}.

\subsubsection{Orientation of the polyhedron}

%

The finite set of measurements $\{ M_{a|x} \}$ we use can be viewed as forming a polyhedron on the Bloch sphere. One may thus wonder whether the orientation of this polyhedron is important.

In the case of an isotropic map, i.e., $\xi = \mathds{1}/2$, changing the orientation of the polyhedron clearly does not change its shrinking factor, while it may change the result of the algorithm. In principle one could thus optimize the algorithm over global rotations of the polyhedron. Nevertheless, we observed this optimization can be avoided when taking into account auxiliary separable states (see next subsection). Specifically, we find that when running the final LHS protocol (see Eq.~\eqref{LHSprotim} below), the optimization over rotations becomes irrelevant, and can be safely omitted. 

On the other hand, for more general maps, which identify a preferred direction on the sphere, the orientation of the polyhedron has in general an impact, and leads to different shrinking factors, which indeed affects the performance of the algorithm. Therefore, in the case of a non-isotropic map one has to first optimize the shrinking factor over all possible rotations of the vertices. Next, one can start from the obtained polyhedron to construct families of polyhedra tailored to a given map, as explained in the next section. 

\subsubsection{Constructing families of polyhedra}

In practice, it is efficient to run the algorithm as a hierarchy of tests. That is, one first chooses a finite set of measurements $\{ M^1_{a|x} \}$ and runs the LHS protocol. This leads to a first value $q^*$ for the visibility of our target state.

Next, one moves to the next level in the hierarchy. Hence we construct a new set of measurements $\{ M^2_{a|x} \}$ with a larger shrinking factor. To do so, we start from $\{ M^1_{a|x} \}$, and identify the facet of the polyhedron that yields the smallest shrinking factor. This allows us to find a specific measurement (leading to this smallest shrinking factor) via Eq.~\eqref{eq:opt_meas}; see Appendix \ref{app:shr2}. Clearly, adding this specific measurement to the new set $\{ M^2_{a|x} \}$ will give rise to a better shrinking factor. Note that several facets may lead to the same smallest shrinking factor, hence the above procedure needs to be repeated until one obtains eventually a strictly better shrinking factor.

The overall process is then repeated until an LHS model is constructed, or the computation becomes infeasible given the accessible resources.

Another option consists in using families of polyhedra, with increasing number of measurements. For instance, starting from a given polyhedron, one can find its geometric dual. One then defines the next polyhedron in the family, as the one given by all vertices of the initial polyhedron and all vertices of the dual. Repeating the process, we obtain a family of polyhedra, with increasing shrinking factors.

\subsection{Adding auxiliary states}

Another way to improve the protocol is by introducing auxiliary states, i.e., adding a list of states that are known to admit a LHS model. Indeed, the condition imposed in \eqref{LHSprot}, $\chi^{\eta} =  \eta \chi + (1-\eta) \xi \otimes \chi_B = \rho_q$, aims at constructing a local state that is equal to the target one $\rho_q$. This condition can be relaxed by demanding that $\rho_q $ equals a \textit{convex combination} of $\chi^{\eta}$ and some other states which admit an LHS model. Clearly, this still implies that $\rho_q$ admits a LHS model.

Formally, the condition \eqref{LHSprot} in the SDP can straightforwardly be generalized to include auxiliary states. Auxiliary states can be chosen to be (i) separable states, (ii) entangled states admitting a LHS model; see e.g.~Refs \cite{Augusiak_review,Bowles14,Jevtic15,Bowles16} for classes of entangled states with LHS models, as well as Ref.~\cite{Cavalcanti16} for a long list of unsteerable states. Another possibility consists in re-using unsteerable states previously obtained from the method.

While the characterization of separable states is in general a difficult problem, the case of two-qubit states stands out as a notable exception, the partial transpose criterion giving a full characterization of separable states \cite{Peres96,Horodecki96}. As the condition of positivity under partial transpose (PPT) can be formulated as an SDP condition, we can exploit this feature in the protocol.

Formally, we rewrite condition \eqref{LHSprot} in the SDP as
\ba \label{eq:sep_state}
\lambda \left( \eta \chi + (1-\eta) \xi \otimes \chi_B \right) + (1-\lambda) \rho_{sep} = \rho_q
\ea
where $\rho_{sep}$ is a separable two-qubit state. This leads to
\ba
\frac{1}{(1-\lambda)} \rho_{sep} =  \rho_q  - (\eta \tilde{\chi} + (1-\eta) \xi \otimes \tilde{\chi}_B )
\ea
which results in the SDP conditions:
\ba
\rho_q  -(\eta \tilde{\chi} + (1-\eta) \xi \otimes \tilde{\chi}_B ) \geq 0 \\
(\rho_q  -(\eta \tilde{\chi} + (1-\eta) \xi \otimes \tilde{\chi}_B ) )^{T_B} \geq 0 \\
\Tr(\tilde{\chi}) \geq 0
\ea
where $\tilde{\chi} = \lambda \chi$, and ${T_B}$ stands for the partial transposition on Bob's side.
Note that the last constraint guarantees that $\lambda \leq 1$.

This idea can also be used for higher-dimensional quantum states. Although the PPT criterion does not guarantee separability anymore \cite{Horodecki97}, one can still check via other means whether the variable $\rho_{sep}$ outputted by the SDP defines indeed a separable state. If this is the case, then the constructed LHS model is valid. Otherwise, one can still try to prove that the output $\rho_{sep}$ is unsteerable. Note that this is however not the case in general, as there exist PPT entangled state that lead to steering \cite{Moroder14} and nonlocality \cite{Vertesi14}.

Moreover, considering entangled states admitting a LHS model may further improve performance. One can then add to Eq.~\eqref{eq:sep_state} a list of unsteerable states. See the final version \eqref{LHSprotim} of the LHS algorithm for details on the implementation.

\subsection{Removing redundant conditions}

When implementing the SDP, it is important to effectively remove all redundant conditions, in order to enhance performance. For a set of $n$-outcome POVMs, normalization allows us to restrict to $n - 1$ outcomes. Given an assemblage $\sigma_{a|x} = \Tr_A(A_{a|x} \otimes \mathds{1}\, \chi)$, with $\sum_{a}A_{a|x} = \mathds{1}$, one has
\ba
\sigma_{N|x} = \chi_B - \sum_{a=1}^{N-1} \sigma_{a|x}\,.
\ea
Therefore, if one has a LHS model for $\sigma_{a|x}$, $a=1,\ldots,N-1$, the elements corresponding to the last outcome can be expressed as
\ba
\sigma_{N|x} &= \chi_B - \sum_{a=1}^{N-1} \sum_\lambda \sigma_\lambda D_\lambda(a|x)  \\ \nonumber
&= \chi_B - \sum_\lambda \sigma_\lambda [1 - D_\lambda(N|x) ]
\ea
and imposing $\sum_\lambda \sigma_\lambda = \chi_B$ one gets
\ba
\sigma_{N|x} &=\sum_\lambda \sigma_\lambda D_\lambda(N|x)
\ea
as required for the LHS model to extend to the elements of the assemblage corresponding to the $n$-th outcome.

One can thus impose only $\sum_\lambda \sigma_\lambda = \chi_B$, and forget about the last outcome, consequently reducing the number of constraints in the SDP by removing overall $m-1$ equations, where $m$ is the number of inputs.

\subsection{Selecting deterministic strategies}\label{sec:strats}

The time required to run the LHS protocol
increases exponentially with the number of measurements in the finite set $ \{ M_{a|x} \}$ considered. This is due to the convex structure of the problem: the SDP essentially finds a decomposition of a point in terms of the deterministic strategies. Considering $m$ measurements with $k$ outcomes, we have $k^m$ deterministic strategies to consider. Hence, as $m$ grows (which is desirable in order to improve the shrinking factor), the problem quickly becomes infeasible.

It is nevertheless possible to circumvent this problem. Instead of considering all deterministic strategies, one can focus on a relatively small subset of them. While the bound we obtain might be suboptimal in general, it will nevertheless hold and allows us to construct an LHS model. Moreover, it turns out that in certain cases, a large subset of the deterministic strategies can be omitted, without loss of generality. Thus, choosing appropriately the subset of deterministic strategies turns out to be important.

When using the protocol sequentially, we found that it is relatively efficient to sort deterministic strategies via an adaptive selection.
Starting with a set containing only few measurements, we run the LHS protocol. The SDP gives back the weights $p_\lambda = \Tr(\sigma_{\lambda})$ associated to each deterministic strategy. One can then remove a subset of strategies which have sufficiently low weights. When moving to the next step of the protocol, the new set of deterministic strategies is generated from only those kept in the previous round. Specifically, each of these strategies leads to a new set of strategies, where the outcome of the additional measurements is added (taking all possibilities into account). Then, we run the SDP. From the output, one can then again sort the relevant deterministic strategies, and move to the next step. Note that after each round, one can check how efficient the selection of the strategies is. Indeed, it suffices to run the SDP again, but using only the selected subset of deterministic strategies. If the result is close to the original one, then the sorting did not affect performance too much. If at the same time, the subset of selected of strategies is small compared to the original one, then the algorithm will be much more efficient in the next round, and allows one to consider finite sets of many more measurements (and thus higher shrinking factors). Finally, note that in practice we found no universal manner of sorting the deterministic strategies. The procedure must be adapted case by case.

\subsubsection{Selection based on a given response function} \label{sign}

It is also possible to base the sorting procedure on a specific response function. For instance, consider the LHS model of Werner \cite{Werner89}. Here, the hidden variable on Alice's side is simply a Bloch vector $\hat{\lambda}$, indicating which qubit state $\sigma_\lambda$ has been sent to Bob. Alice now receives a measurement direction $\hat{v}$ and outputs $a=\pm 1$ with probability
\ba
p(\pm|\hat{v},\hat{\lambda}) = \frac{1 \pm \text{sign}(\hat{v} \cdot \hat{\lambda} ) }{2} .
\ea
That is Alice outputs $+1$ whenever the measurement vector $\hat{v}$ is in the half sphere around $\hat{\lambda}$, and $-1$ otherwise.

Now, when running the protocol, we are given a set of $m$ measurements, with vectors $ \hat{v}_1,\hat{v}_2,\ldots,\hat{v}_m $. Choosing Werner's response function, we restrict to those deterministic strategies that are compatible with it, which is indeed a strict subset in general. For instance, given three vectors which are not in the same half sphere, one  cannot always get the same outcome. In this way, a large fraction of deterministic strategies can be eliminated. On a standard computer, we can go up to $m \sim 200$ measurements, leading to high shrinking factors: $\eta^* \simeq 0.99$. This is tremendous progress compared to the case where one would have to keep all $2^m$ deterministic strategies; here the problem would only be feasible up to $m=16$.

Note that, more generally, we observe that Werner's response function appears to be optimal whenever $\xi = \mathds{1} /2$ (or when $\Tr_A(\rho_{sep})=\mathds{1}/2 $).

\subsection{Algorithm: final version}\label{sec:final_version}

We are now ready to provide a final version of the algorithm. Again we focus here on the case of LHS models, but a similar protocol for LHV models is given in Appendix \ref{app:lhv}. The algorithm is particularly tailored to the case of two-qubit states.

Given a target state $\rho$, we consider the family of states $\rho_q = q \rho + (1-q) \rho_{sep}$. Our goal is to determine the largest visibility $q$ such that $\rho_{q}$ admits a LHS model. Ideally, we find $q=1$, in which case $\rho$ is unsteerable.

We first define the noise map $\Phi^{\eta}$ in Eq.~\eqref{map} by choosing the density matrix $\xi$. As discussed above, setting $\xi = \Tr_B (\rho_{sep})$ appears to be the best choice. Next one chooses a finite set of measurements $\{ M_{a|x} \}$ and compute its shrinking factor $\eta$ (which depends on $\xi$). One can then run the following SDP:

\medbreak

{\bf LHS Protocol (final version)}
\begin{align} \label{LHSprotim} \text{find  } & q^* = \max  q  \\
  \text{s.t.  }   & \Tr_A( M_{a|x} \otimes \mathds{1} \, \chi) = \sum_\lambda \sigma_\lambda D_\lambda(a|x) \quad \forall a,x   \nonumber\\
  &  \rho_q - \eta \chi + (1-\eta) \xi \otimes \chi_B -\sum_k p_k \, \rho_{lhs}^k \geq 0 \nonumber\\
  &  \left( \rho_q - \eta \chi + (1-\eta) \xi \otimes \chi_B -  \sum_k p_k \,\rho_{lhs}^k \right)^{T_B}\!\! \geq 0 \nonumber\\
  & \Tr(\chi) + \sum_k p_k \, \Tr(\rho_{lhs}^k ) \geq 0 \nonumber\\
  &  \sigma_\lambda \geq 0\quad \forall \lambda, \quad  p_k \geq 0\quad \forall k.\nonumber
\end{align}
The SDP variables are (i) the positive matrices $\sigma_\lambda$ and positive coefficients $p_k$ and (ii) a hermitian matrix $\chi$. Also, $\{\rho_{lhs}^k\}$ is a list of states admitting a LHS model. The index $\lambda = 1,\ldots,n$ runs over deterministic strategies $D_\lambda(a|x)$. As discussed above, it can be advantageous in practice to restrict to a well-chosen subset of deterministic strategies, which can considerably speed up the SDP without affecting the result (i.e., returning a value of $q^*$ which is essentially the same as if considering all strategies).

At this point, the first level has been completed, with a resulting visibility $q^*$. One can then move to the next level as follows. The idea is to consider a new finite set of measurements $\{ M^2_{a|x} \}$, featuring more measurements than the one used in the first level. For instance, we discussed above how to efficiently construct $\{ M^2_{a|x} \}$ by complementing the initial set. This results in a new shrinking factor $\eta_2$, which is equal or greater than $\eta$ (note that $\xi$ is the same as in the first level). With these parameters, the SDP can be run again. As the set $\{ M^2_{a|x} \}$ features now more measurements, there are in principle more deterministic strategies to be considered. As discussed above, there are several options for efficiently selecting the deterministic strategies, in order to limit the number of SDP variables.

\begin{center}
  \begin{figure} [t!]
    \includegraphics[scale=0.4]{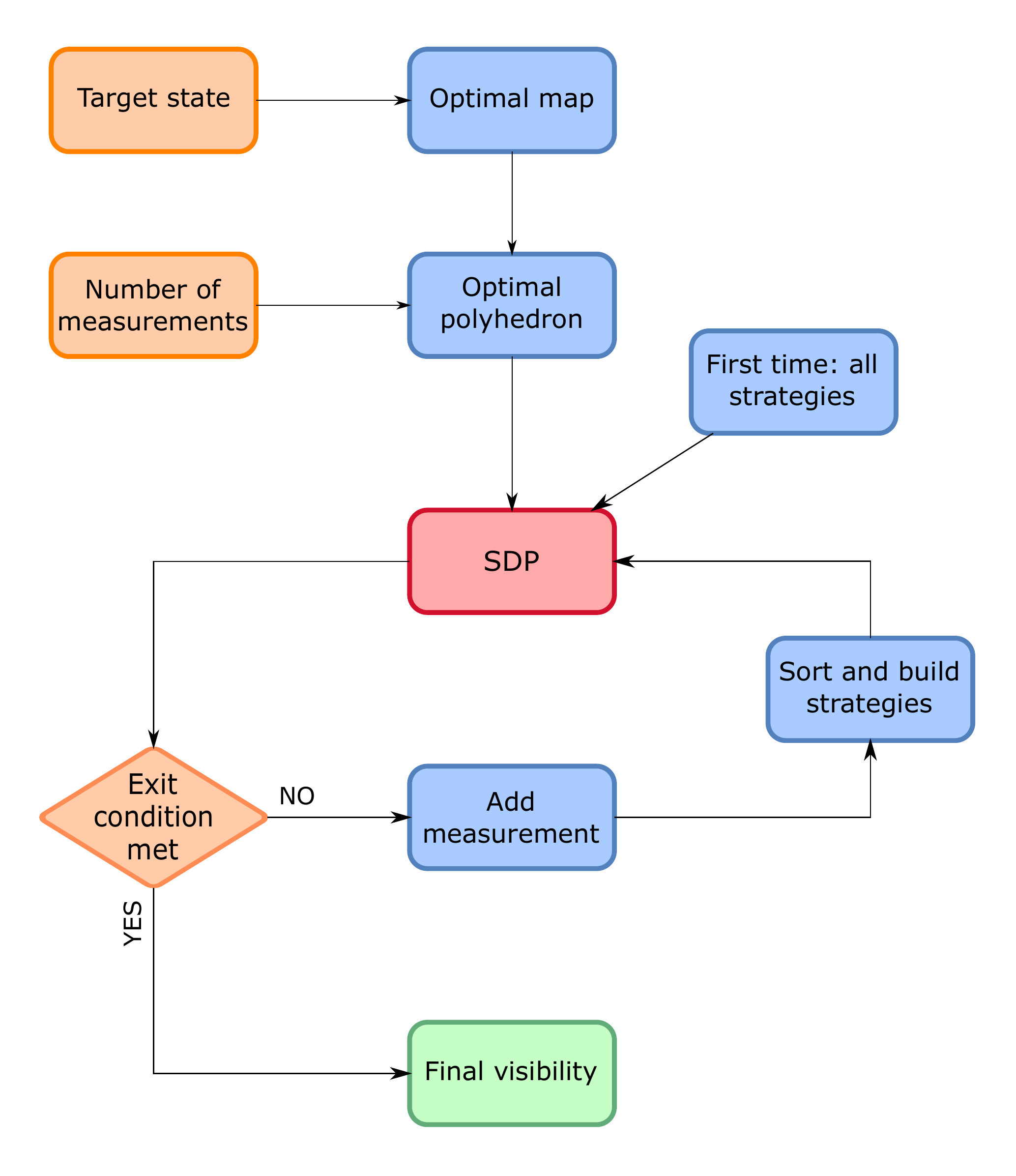}
    \caption{Schematic view of the final algorithmic method.}
    \label{Fig_structure}
  \end{figure}
\end{center}

At this point, the second level has been completed, resulting in a visibility $q^*_2 \geq q^*$.\footnote{Note that the visibility will not decrease using the iterative construction for the finite set of measurements. In general, however, it can be that the visibility decreases when moving to a higher level of the protocol.} Then, the procedure can be repeated as long as computational resources allow for it. The general structure is sketched in Fig.~\ref{Fig_structure}.

In practice, starting from sets of 6 to 10 measurements, it is possible to reach sets of more than 100 measurements, when efficiently selecting deterministic strategies. Examples will be discussed in the next section. More generally, note that the algorithm is proven to converge in the limit, i.e., if $\rho$ admits a LHS model, the algorithm will in principle find it.\footnote{However, note that depending on the method used to select the deterministic strategies, the algorithm might not converge in the limit.}

\section{Analytical solutions} \label{analytical}

One could argue that, because of numerical precision limit, all the results obtained using the algorithm are not sufficient to rigorously guarantee the existence of a local model. In this section, we prove that computer imprecision can be cured to get an analytical local model from the one obtained numerically. We discuss the case of LHV models as this makes the analysis simpler (see Appendix \ref{app:lhv} for a discussion of the algorithm for LHV models). The case of LHS models can be treated with similar ideas.

The first step is to tighten the inequality constraints in \eqref{LHVprot} to make them tolerant to this imprecision.
Specifically, we change the constraint $p_\lambda\geq0$ into $p_\lambda\geq\epsilon$ where $\epsilon$ is the numerical precision. Hence, even if the computer overestimates the ``actual'' value of $p_\lambda$, this value is ensured to remained positive.

The second step concerns the equality constraint in \eqref{LHVprot} and is less trivial.
Because of numerical imprecision, $p(ab|xy)=\sum_\lambda p_\lambda D_\lambda(ab|xy)$ is not satisfied exactly but becomes
\begin{equation}
  p(ab|xy)=\sum_\lambda p_\lambda D_\lambda(ab|xy)+r_{ab|xy}
  \label{eqn:lhvana1}
\end{equation}
with $|r_{ab|xy}|\leq\epsilon$. Note that, due to numerical imprecision, $\sum_\lambda p_\lambda D_\lambda(ab|xy)$ is ill-normalized.

The idea is to show that, for small enough $\epsilon$, $r_{ab|xy}$ always admits a local model, i.e. it can be decomposed as $\sum_\lambda r_\lambda D_\lambda(ab|xy)$ with $r_\lambda\geq0$.
If this holds, then $p(ab|xy)=\sum_\lambda (p_\lambda + r_\lambda) D_\lambda(ab|xy)$ with $p_\lambda+r_\lambda$ positive coefficients summing up to one by normalization of $p(ab|xy)$.
We first give geometrical insight to this problem and then solve it more formally in the two-outcome scenario.

Following Refs \cite{Rosset14,Renou17}, we split Eq.~\eqref{eqn:lhvana1} into its normalization and no-signaling (NS) parts. The convenient feature of this representation is that the maximally mixed distribution $p_0(ab|xy) = 1 / N_O^2$ for all $a,b,x,y$, where $N_O$ denotes the number of outputs, is located at the origin. 
While the normalization part is essentially trivial, the no-signaling reads
\begin{equation}
  p^{\mathrm{NS}}(ab|xy)=\sum_\lambda p_\lambda D_\lambda^{\mathrm{NS}}(ab|xy)+r_{ab|xy}^{\mathrm{NS}}.
\end{equation}
In the no-signaling vector space, $r_{ab|xy}^{\mathrm{NS}}$ is in the ball of radius $\epsilon$ centered around the origin.
If this ball is contained in the local polytope, i.e., the convex hull of all deterministic strategies, then we can conclude.
Intuitively, it can be seen immediately that this is the case. This is because the origin, i.e., the distribution $p_0$, can be viewed as the ``center'' of the local polytope, obtained e.g., by an equal mixture of all deterministic local strategies. 

Next, one needs to estimate out how large $\epsilon$ can be such that the ball is still inside the local polytope. In order to do so, one can find the radius of the largest ball (centered at the origin) that can fit inside the local polytope. Equivalently, one should find a point on the surface of the local polytope that is closest to the origin. While we could not derive a general solution here, we nevertheless conjecture that the closest point is always on a positivity facet (and not on a Bell inequality), so that its euclidean distance to the origin is $1/N_O^2$. We verified this for the case of binary and ternary inputs and binary outputs.

In the case of binary outcomes one can get an explicit condition on $\epsilon$ guaranteeing that $r_{ab|xy}$ is local. This is an extension of a procedure presented in \cite{Hirsch17}. First, we transform $r_{ab|xy}$ to the no-signaling representation in terms of the correlators $C_{xy} = p(a=b |xy) - p(a \neq b |xy)$, and the local marginals $C_{x0} = p(a=+1 |x)-p(a=-1 |x)$ and $C_{0y} = p(b=+1 |y)-p(a=-1 |y)$. We obtain a matrix $C$, with coefficients $C_{ij}$, which can be decomposed as follows (note that the coefficient $C_{00}$ is irrelevant)
\ba \label{loc_decomp}
C = \sum_{ij} \abs{C_{ij}} \text{sign}(C_{ij}) T_{ij}
\ea
where $i,j \in \{ 0,\ldots, N_I\}$ with $N_I$ denoting the number of inputs. Here, $T_{ij}$ is the matrix having entry $+1$ at position ($i,j$) and zeros elsewhere. These represent local distributions. For $i,j>0$, they can be obtained by random outputs for all inputs except for input $i$ for Alice and input $j$ for Bob, for which they perfectly correlated their output (i.e., both output $+1$ or $-1$ with probability $1/2$, resulting in random marginals but a correlated joint outcome). The remaining terms $T_{i0}$ (and $T_{0j}$) are obtained by Alice outputting $+1$ for measurement $i$ (measurement $j$ for Bob, respectively) and randomly for all other measurements. Finally, the outputs can be flipped or not depending on the coefficient $\text{sign}(C_{ij})$. Thus, we conclude that $r_{ab|xy}$ is local whenever
\ba
\sum_{ij} \abs{C_{ij}} \leq 1 .
\ea
This provides a simple condition (far from being optimal however) ensuring that $r_{ab|xy}$ is local given it is close enough to the origin.

\section{Results} \label{results}

We now apply our ready-to-use algorithm for constructing LHS models for several families of entangled two-qubit states, considering projective measurements. In particular we consider the case of Bell diagonal states, for which the optimal LHS model has been recently obtained \cite{Jevtic15,Nguyen16,Zhang17,Yu17}. This allows us to benchmark our algorithm against the exact solution. Notably, we find that our method construct LHS models that are close to optimal, which is a good indication that it works very efficiently for two-qubit states. Moreover, we discuss other families of states, where we compare our models to the best-known bounds on steerability, and find that they are relatively close in general.

Generally, we observe that our systematic implementation of the algorithm provides relevant results for all investigated classes of states, within a reasonable amount of computation time (up to one day per target state).

\subsection{Bell diagonal states}

Bell diagonal states are convex combinations of Bell states, and can be written in the Pauli basis as follows
\begin{equation}\label{BD}
  \rho_{BD}=\frac{1}{4}\left(\mathds{1}\otimes\mathds{1} + \sum_{i=1}^3t_{i}\sigma_{i}\otimes\sigma_{i}\right)
\end{equation}
where $\vec{t} \in \mathds{R}^{3}$ is the correlation vector. As mentioned above, an optimal LHS model was recently derived. First presented and conjectured to be optimal in \cite{Jevtic15}, the model was later proven to be optimal in \cite{Nguyen16}; more recently two alternative models were proposed \cite{Zhang17,Yu17}. More specifically, these results provide a criterion, based on the norm of correlation vector components $s_i = |t_i|$, which exactly separates steerable from unsteerable Bell diagonal states. Unfortunately, this cannot be expressed in simple form, so we refer the reader to Refs \cite{Jevtic15,Nguyen16,Zhang17,Yu17} for details.

\begin{figure} [b!]
  \centering
  \includegraphics[width=\columnwidth]{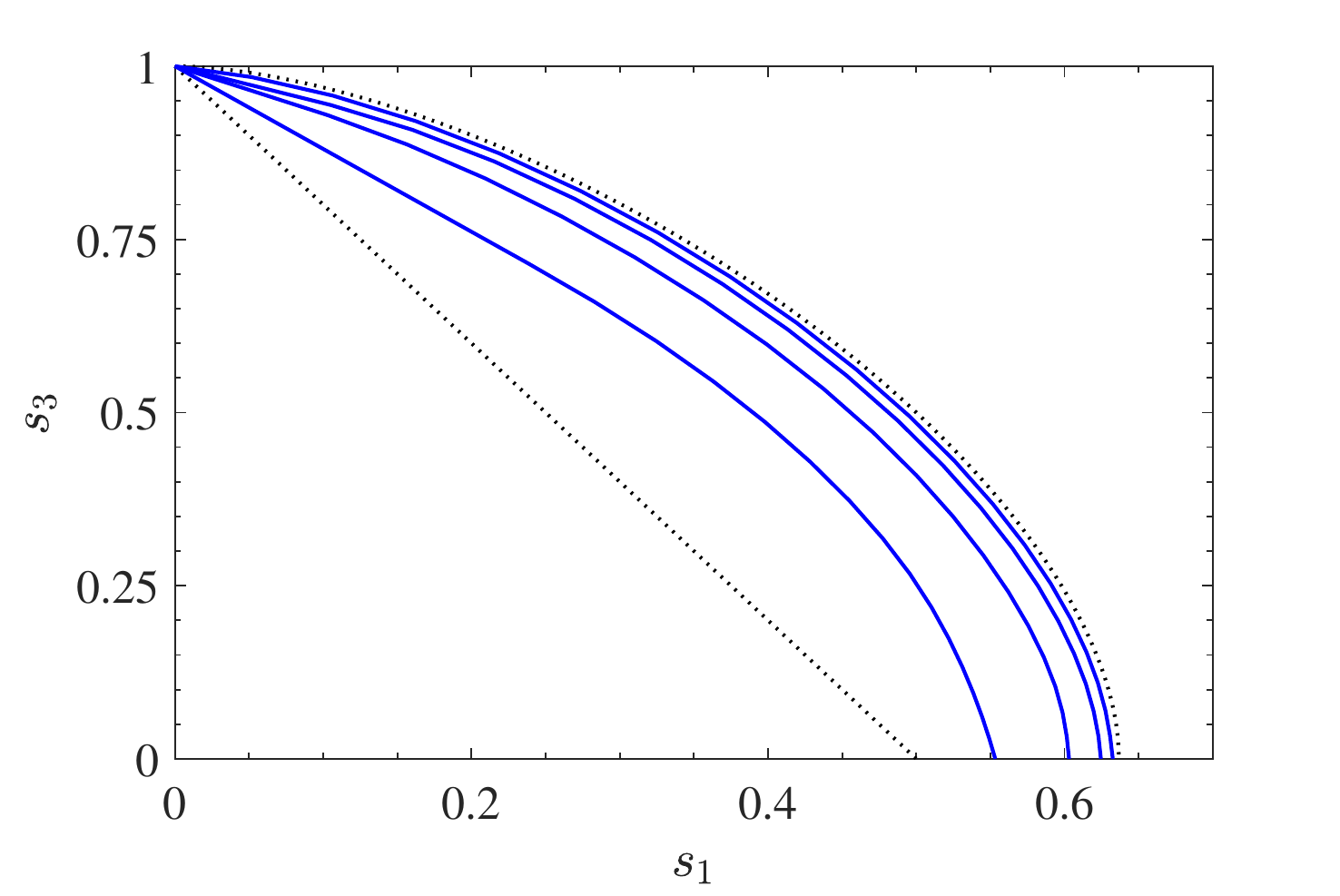}
  \caption{Steering properties of Bell diagonal states with $s_1=s_2$. The states are separable below the dotted line, while they admit a LHS model below the dotted curve; this represents the exact steering limit as shown in Refs. \cite{Jevtic15,Nguyen16}. Using our LHS protocol, we obtain the four solid curves, corresponding to each level. At level four the difference with respect to the exact steerability limit is less than 1\%, illustrating the efficiency of our method in this case.}\label{Fig_BD}
\end{figure}

We first investigate the case $s_1=s_2$, and construct LHS models using our algorithm. The results can be conveniently represented in the plane $(s_1, s_3)$, as shown in Fig.~\ref{Fig_BD}. We see that the constructed models quickly approach the steerability limit, and are thus very close to optimal.

In practice the procedure is implemented as follows. We first define $\rho_q$ by choosing $\rho = \rho_{BD}$ with $s_3 = 1$ and varying $s_1 \in [0,1]$, and $\rho_{sep}=  \mathds{1}/4$. Given the form of the states, we set $\xi  = \mathds{1}/2$. In this first iteration, we consider a set of six projective measurements. Here, the optimal choice is to have measurements such that their Bloch vectors form an icosahedron on the sphere; this achieves the highest shrinking factor $\eta = \sqrt{(5+2\sqrt{5})/15} \simeq 0.7947$. Then, we run the protocol by increasing the number of measurements up to 136. In each step, we select the relevant deterministic strategies, that is, we sort them by weights and find the smallest $n$ such that the first $n$ deterministic strategies suffices to find the same answer when running the protocol again (in practice, one can run the protocol using the first strategies, then the first two, and so on until one gets the desired result, up to the SDP precision). We observe that this procedure naturally leads to four distinct levels, each of which corresponds to an increase of the shrinking factor. Specifically, the levels are given by 6, 36, 96, and 136 measurements, with corresponding shrinking factors $\eta \simeq 0.79$, $0.92$, $0.97$, and $0.99$. At each new level, the polyhedron constructed is the one of the previous level, together with its geometric dual. Fig.~\ref{Fig_BD} shows how larger classes of states are detected at each successive level. 

It should be pointed out that these results can also be obtained via a slightly different method. We start from the icosahedron, and define a recursive family of polyhedra by taking the previous polyhedron and add its geometric dual. Considering the first four polyhedra in this family, we run the LHS protocol, selecting the deterministic strategies that are compatible with the sign response function (see Section \ref{sign}). This leads to four curves, which are equivalent to those generated above. The running time of both methods is similar. Hence, we conclude that our first method, which is more systematic, performs actually well in practice.

\subsubsection{Rank-3 entangled states}\label{rank3}

Consider mixtures of three Bell states (hence Bell diagonal of rank 3) of the form
\begin{equation}
  p_{1}\ketbra{\psi^{-}} + p_{2}\ketbra{\psi^{+}} + (1{-}p_{1}{-}p_{2})\ketbra{\phi^{+}}. \nonumber
\end{equation}
These states are interesting to consider, as they are not full rank, i.e., they lie on the border of the Hilbert space.
Hence, when running the LHS protocol, the SDP variable $\chi$ must lie outside the Hilbert space (i.e., $\chi$ is no longer positive semi-definite), in order for the ``shrunk'' state $\chi^\eta$ to be mapped on the border of the Hilbert space. We will see that this is indeed the case, and that our method works well even when considering non-full-rank entangled states.

\begin{figure}
  \centering
  \includegraphics[width=\linewidth]{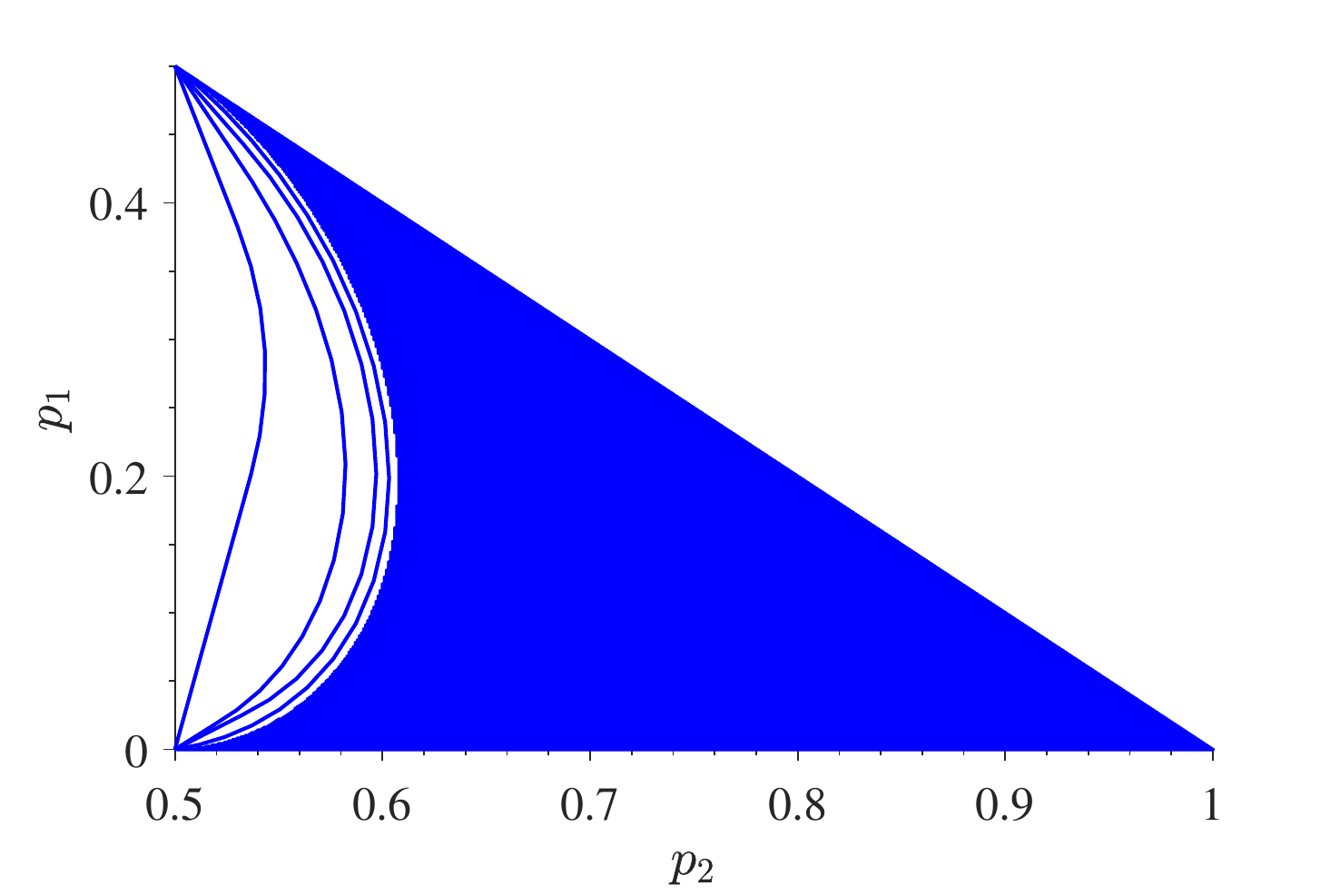}
  \caption{Steering properties of rank-3 Bell diagonal states. States are entangled in the entire region. They are unsteerable inside the white region, while states in the blue region are steerable. Implementing the protocol from level one to four, we obtain the four curves. Again, we observe that level four is very close to being optimal.}
  \label{Fig_rank3}
\end{figure}

By symmetry, it is enough to focus on the region $p_1 \in [0,1/2]$ and $p_2 \in [1/2,1]$, where the states are entangled (as checked, e.g., via partial transposition). As above, we run the protocol starting from $6$ measurements, and up to 136 measurements. This again defines four distinct levels. The results are given in Fig.~\ref{Fig_rank3}. Again, we observe that the method quickly converges to the exact steerability limit, and that the curve obtained at level four is extremely close to being optimal.

\subsection{Entangled states with non-uniform marginals}

Next, we investigate the performance of our LHS protocol on different classes of entangled states, featuring reduced states that are not maximally mixed. Benchmarking the performance of our protocol is more complicated in this case, as the exact steerability limit is not known. Nevertheless, we can use known bounds on the steerability of these states in order to get an estimate. Again, we observe that the performance of our protocol is good, and that the LHS models it constructs are generally close to being optimal. Moreover, these results show that the sufficient condition for unsteerability presented in Ref.~\cite{Bowles15} is in general not necessary.

\subsubsection{Partially entangled states with white noise}

\begin{figure}
  \includegraphics[width=\linewidth]{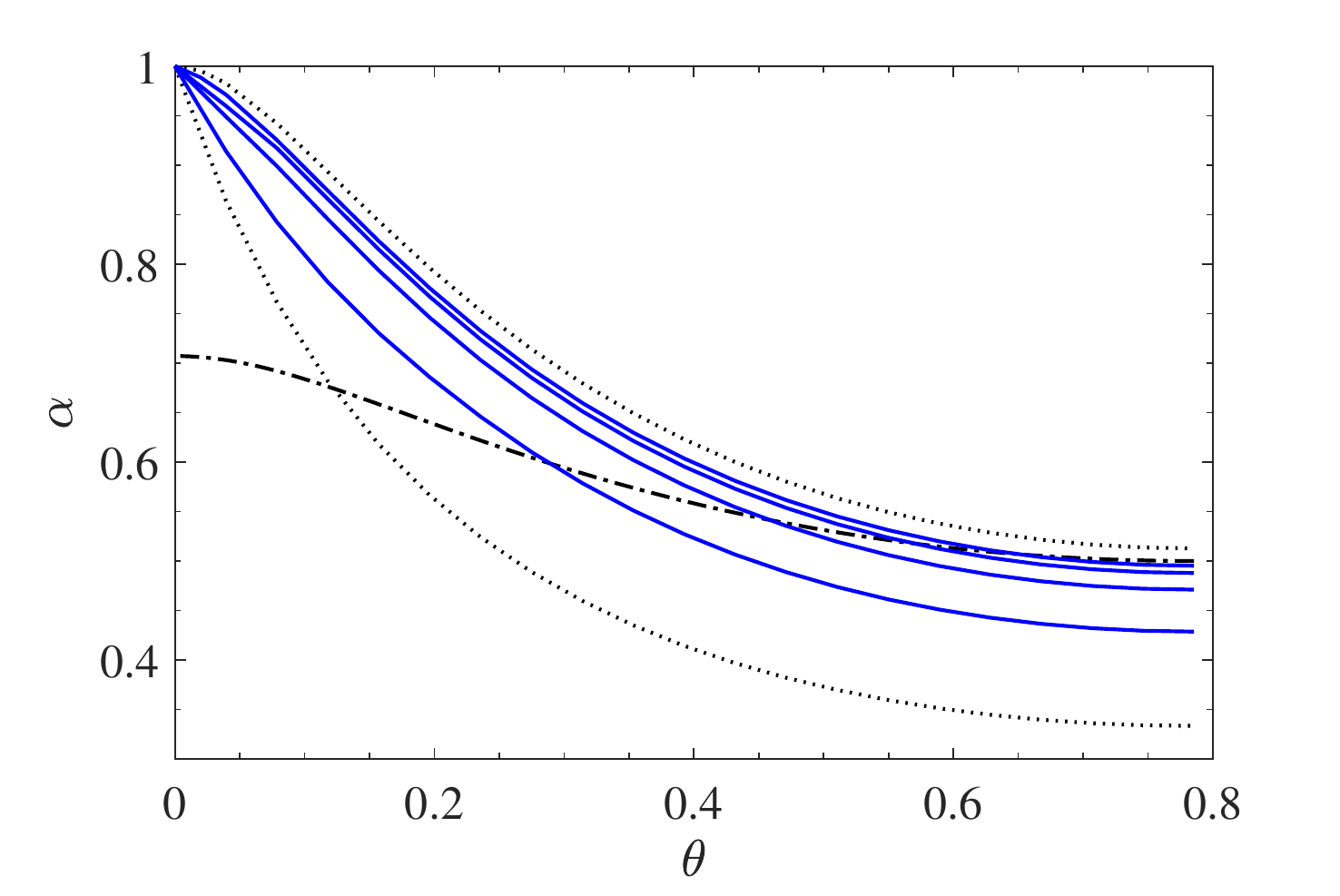}
  \caption{Steering properties of partially entangled states mixed with white noise, see Eq.~\eqref{MMM}. The lower dotted curve is the separability limit: states below this curve are separable. The upper dotted curve is a sufficient condition for steering: any state above this curve is steerable, as found via SDP techniques \cite{paul_review} (using 13 measurements on the Bloch sphere). The four solid curves represent the results for of the protocol from level one to four, with growing number of measurements from bottom to top (6, 16, 46, and 136 measurements). The dashed-dotted curve corresponds to a sufficient condition for a state to be unsteerable: states below this curve admit a LHS model \cite{Bowles15}. Our results show that this criterion is not tight in general.}
  \label{Fig_MMM}
\end{figure}

We consider a natural extension of Werner states, replacing the maximally entangled state by a partially entangled one

\begin{equation} \label{MMM}
  \rho(\alpha,\theta) = \alpha\ket{\psi_{\theta}}\bra{\psi_{\theta}} + (1{-}\alpha)\mathds{1}/4
\end{equation}
where $\ket{\psi_{\theta}}=\cos\theta\ket{00}+\sin\theta\ket{11}$.

We run the LHS protocol for different values of $\theta$, aiming at maximizing the visibility $\alpha$. In all cases, we set $\xi= \mathds{1}/2$. We start from six projective measurements; as $\xi= \mathds{1}/2$, the best polyhedron is still the icosahedron. As in the case of Bell diagonal states, we observe here that two methods give similar results. The first option consists in increasing the measurement number up to 136, selecting the deterministic strategies in each step. This identifies four levels, where the shrinking factor increases. The second option is to consider the family of 4 polyhedra (generated from the icosahedron, and adding the geometric dual in each step), and use the sign response function to select deterministic strategies. As explained in section \ref{sign}, sorting the deterministic vertices following Werner's sign function seems to be optimal\footnote{It is probably \textit{over-optimal}, in the sense that it selects too many deterministic strategies.} when we set $\xi = \mathds{1}/2$, the curves obtained in both cases are thus similar.

The results are present in Fig.~\ref{Fig_MMM}. We believe that the curve obtained at level four (136 measurements) is close to the steerability limit, that is, closer to the critical curve than the dotted blue curve, representing an upper bound that we derived numerically. We note that, while these results are similar for this class of states to those obtained in \cite{Hirsch16}, our systematic implementation could reproduce these results in a smaller amount of time.

\subsubsection{Partially entangled states with colored noise} \label{resM}

Next, we consider a different class of states, of the form
\ba \label{Mafalgo}
\rho'(\alpha,\theta) = \alpha \ket{\psi_\theta} \bra{\psi_\theta} + (1 - \alpha )  \rho_A \otimes \frac{\mathds{1}}{2}
\ea
where $\rho_A = \Tr_B(\ket{\psi_\theta} \bra{\psi_\theta} )$. These states can be obtained by applying a local filter (on Bob's side) on Werner states \cite{Almeida07}. They are entangled when $\alpha > 1/2$, and separable otherwise. A sufficient criterion for these states to admit a LHS model has been derived in \cite{Bowles15}. Namely, $\rho'(\alpha,\theta)$ is unsteerable as long as
\ba \label{condj}
\cos^2{2\theta}  \geq \frac{2\alpha-1}{(2-\alpha)\alpha^3} .
\ea
Here, we apply the protocol, fixing different values of $\theta$ and maximizing the visibility $\alpha$. Note that in this case, given the form of the state, we do not use an isotropic map anymore, and set $\xi =\rho_A $. We start again with $6$ measurements. However, the optimal set does not form an icosahedron anymore, as the map is not isotropic.

\begin{figure}
  \centering
  \includegraphics[width=\linewidth]{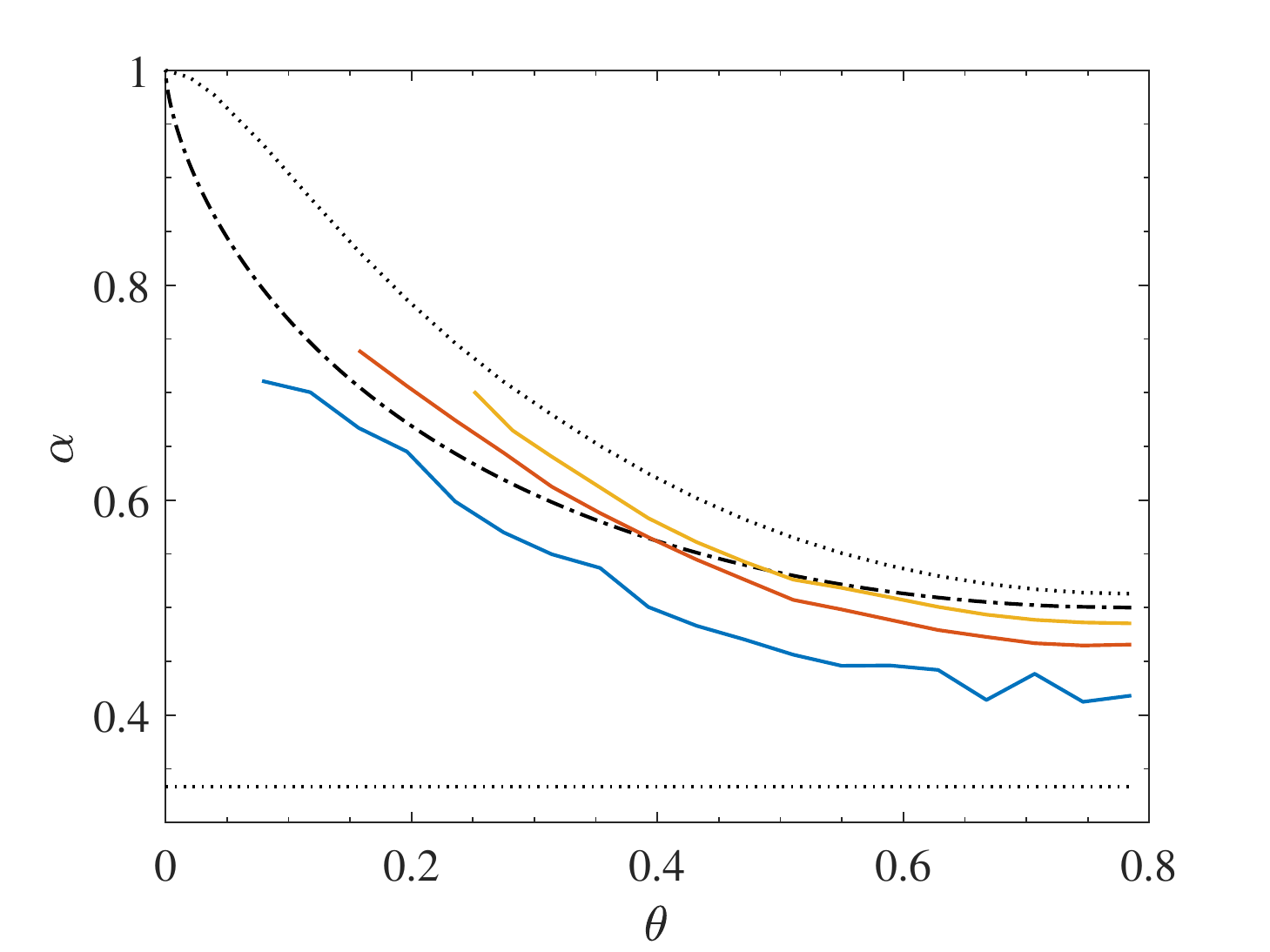}
  \caption{Steering properties of states \eqref{Mafalgo}. States below the lower dashed line are separable, while the upper dashed black curve is an upper bound for steerability found numerically (via SDP techniques \cite{paul_review} and using 13 measurements on the Bloch sphere). The dashed-dotted curve correspond to the best known LHS models so far, i.e., given by Eq.~\eqref{condj}. The solid curves represent three successive levels of the algorithm (corresponding to fixed shrinking factors $\eta_1 = 0.79$, $\eta_2 = 0.92$, and $\eta_3 = 0.97$). This shows that the criterion \eqref{condj} of Ref.~\cite{Bowles15} is not tight.}
  \label{Fig_Maf}
\end{figure}

We ran three levels of the hierarchy, each level corresponding to a fixed value of $\eta$. We chose $\eta_1 = 0.79$, $\eta_2 = 0.92$, and $\eta_3 = 0.97$, corresponding to the first three levels discussed above in the case of $\xi= \mathds{1}/2$. Our results are given in Fig.~\ref{Fig_Maf}. They show explicitly that the criterion of Ref.~\cite{Bowles15} is in general not necessary, as we obtain better LHS models. Note that running the protocol at level three is computationally demanding, especially for small values of $\theta$, as the number of measurements required to reach a given shrinking factor $\eta$ increases with the purity of $\xi  = \rho_A $. Hence, the number of measurements required for $\eta_3$ becomes very large which is the reason why we did not run any point at level three in this regime.

Finally, we note that the choice of a non-isotropic noise map is here important. We checked that when using the isotropic map, the obtained curves are much weaker. While they would give similar results for $\theta = \pi/4$, the visibility $\alpha$ then decreases when $\theta$ decreases.

\section{Conclusion}\label{conclusion}

We have discussed in detail the implementation of an algorithmic procedure for constructing local models for entangled states. Focusing on the case of LHS models for two-qubit states and projective measurements, we showed how each parameter involved in the protocol can be adjusted in order to enhance performance. Moreover, we applied our ready-to-use algorithm to different classes of entangled two-qubits states. First, discussing steering with Bell diagonal states, for which the exact steerability limit is known, we could benchmark the efficiency of our method, and found that the constructed LHS models recover almost the steerability limit. Then we showed that the method also works efficiently for other classes of entangled states. Overall, these results show that our protocol allows one to construct an LHS model for most two-qubit states that admit one. 

Finally we discuss a number of interesting open questions. First, it would be good to understand how to efficiently implement the method for POVMs (e.g.~starting with qubits), as well as for higher-dimensional systems. Here, the main difficulty is to constructs finite sets of few measurements that approximate well the entire set of POVMs (or projective measurements), thus leading to a shrinking factor that is relatively large. An interesting open problem which the method may help addressing, is whether POVMs provide an advantage over projective measurements for demonstrating steering or nonlocality of entangled states \cite{Nguyen17}.

Already for systems of two qutrits there are many interesting questions that could by tackled using our methods. For instance, while it is known that the fully anti-symmetric states leads to strong steering \cite{Paul14}, it is not known whether it can violate a Bell inequality. If this is not the case, then the method could help constructing an LHV model for it. Also, entangled two-qutrit states exhibit an effect of ``anomaly of nonlocality'', in the sense that less entangled states can lead to larger Bell inequality violations \cite{Acin02}. It would be interesting to see if such an anomaly is also present when constructing local models for noisy version of these states.

Finally, the method could also help to construct local models assisted with some nonlocal resource, for instance, classical communication or nonlocal boxes. Here, a long-standing problem is whether one bit of classical communication is enough to simulate the correlations of all entangled pure two-qubit states. While this is proven for the maximally entangled state \cite{Toner03}, the case of partially entangled states is still open. Considering nonlocal boxes, it is known that the maximally entangled state can be simulated with one Popescu-Rohrlich \cite{PR} (PR) nonlocal box \cite{Cerf05}, while very weakly entangled states require at least two PR boxes \cite{Brunner05}. What about strongly but not maximally entangled ones?

\emph{Acknowledgements.} We thank Joe Bowles for discussions. This work was supported by the Swiss national science foundation (starting grant DIAQ and QSIT).

\begin{appendix}

  \section{Computation of the shrinking factor for qubit two-outcome measurements}
  \label{app:shr2}

  Consider the set of projective measurements on qubits, $\mathcal{M}$, with POVMs
  \begin{equation}
    A_{\pm} = \frac{\mathds{1} \pm \hat{v} \cdot \vec{\sigma}}{2}
  \end{equation}
  for some normalized $\hat{v} \in \mathds{R}^3$.

  Using the noise map defined in \eqref{map} the POVM elements of the noisy set $\mathcal{M}^{\eta}$ read as
  \begin{equation}
    A^{\eta} = \left(\frac{1}{2} + (1{-}\eta)\frac{\vec{u}\cdot\hat{v}}{2}\right)\mathds{1} + \frac{\eta\hat{v}\cdot\vec{\sigma}}{2}
  \end{equation}
  where $\xi_{A} = \left(\mathds{1} + \vec{u}\cdot\vec{\sigma}\right)/2$.

  These measurements are thus characterized by the four-dimensional vector
  \begin{equation}
    \textbf{v}^{\eta} = \left\lbrace\left(\frac{1}{2} + (1{-}\eta)\frac{\vec{u}\cdot\hat{v}}{2}\right),\frac{\eta \hat{v}}{2}\right\rbrace
  \end{equation}
  in the four-dimensional space spanned by $\left\lbrace\mathds{1},\vec{\sigma}\right\rbrace$.

  Note that when $\xi_{A} = \mathds{1}/2$, the first component of this vector is constant and always equal to $1/2$ and can thus be ignored. Without loss of generality, the problem is then reduced to a three-dimensional problem, which is why the Bloch representation is sufficient in this case.

  Now, using the facets representation, a point $\textbf{p} ~\in~ \mathds{R}^{4}$ is inside a polytope if and only if
  \begin{equation}
    \left(\textbf{F}_{j},\textbf{p}\right) \leq b_{j} \qquad \forall j = 1,\ldots,N_{F},
  \end{equation}
  where $\left\lbrace\textbf{F}_{j}\right\rbrace_{j = 1}^{N_{F}} ~\in~ \mathds{R}^{4}$ are the facets of the polytope, with bounds $\left\lbrace b_{j}\right\rbrace_{j=1}^{N_{F}}~\in~ \mathds{R}$ and $(~,~)$ is the dot product.

  For the problem we consider, dropping the $j$ index for clarity, the shrinking factor associated to a facet $\textbf{F}$ is the largest value $\eta^{*}$ such that
  \begin{equation}\label{eq:facets}
    \left(\textbf{F},\textbf{v}^{\eta}\right) \leq b \qquad \forall \; \textbf{v}^{\eta} \in \mathcal{M}^{\eta} .
  \end{equation}
  For this value $\eta^{*}$, there is only one vector $\textbf{v}^{\eta^{*}}_{m}$ that saturates the inequality,

  \begin{equation}\label{eq:vec_satur}
    \left(\textbf{F},\textbf{v}^{\eta^{*}}_{m}\right) = b.
  \end{equation}
  However, it is easier to find and solve the dual problem of Eq.~\eqref{eq:vec_satur}, i.e.
  \begin{equation}\label{eq:opt_meas_facet}
    \left(\textbf{F}^{\eta^{*}},\textbf{v}_{m}^{1}\right) = b,
  \end{equation}
  where $\textbf{v}^{1}$ is a vector of the form $\frac{1}{2}\left\lbrace 1, \hat{v}\right\rbrace$, and the new facet vector $\textbf{F}^{\eta}$ is defined as
  \begin{equation}
    \textbf{F}^{\eta} = \left\lbrace F_{0},\eta\hat{F} + (1{-}\eta)\vec{u}F_{0}\right\rbrace \equiv \left\lbrace F_{0},\vec{n}(\eta)\right\rbrace.
  \end{equation}
  where we have defined $\textbf{F} = \left\lbrace F_{0},\hat{F}\right\rbrace$.

  \medbreak

  It is then easy to see that, in order to maximize the scalar product given by Eq.~\eqref{eq:opt_meas_facet} and actually saturate the inequality, $\textbf{v}^{1}$ has to be of the form
  \begin{equation}\label{eq:opt_meas}
    \textbf{v}^{1}_{m} = \frac{1}{2}\left\lbrace 1,\frac{\vec{n}(\eta^{*})}{\norm{\vec{n}(\eta^{*})}}\right\rbrace.
  \end{equation}
  One can then invert that formula to obtain a quadratic expression for $\eta(\vec{F},b,\xi_{A})$, namely,
  \begin{equation}
    A\eta^{2} + B\eta + C = 0,
  \end{equation}
  with
  \begin{equation}
    \begin{split}
      A & = \sum_{k=1}^{3}\left(F_{k} {-} F_{0} u_{k}\right)^{2}\\
      B & = 4F_{0}\sum_{k=1}^{3}u_{k}\left(F_{k} {-} F_{0}u_{k}\right)\\
      C & = 4b\left(F_{0} {-} b\right),
    \end{split}
  \end{equation}
  $\eta^{*}$ being the largest of the two solutions to this equation.
  \medbreak

  The shrinking factor of a polyhedron is then simply given by
  \begin{equation}
    \eta^{*}\left(\left\lbrace A_{a|x}\right\rbrace,\xi_{A}\right) = \underset{j}{\text{min}} ~\eta^{*}_{j}\left(\vec{F}_{j},b_{j},\xi_{A}\right), \qquad j = 1,\ldots,N_{F}.
  \end{equation}
  This way, we ensure that any element of the continuous set $\mathcal{M}^{\eta}$ fulfills condition \eqref{eq:facets}, while obtaining the largest possible value for the shrinking factor $\eta^{*}$.

  \section{Protocols for LHV models} \label{app:lhv}

  First, we define the following noisy map:
  \begin{equation}
    \Phi^{\eta}(M_{a})  = \eta M_{a} + (1-\eta) \Tr(\xi_{a}) \mathds{1} \equiv M_{a}^{\eta}
  \end{equation}
  where $0 \leq \eta \leq 1$ and $\xi_a$ is a density matrix. Note that $\Phi^{\eta}$ maps valid POVMs into valid POVMs, when applied to each POVM element of the set. Next, it is easy to see that the statistic of these noisy measurements on a state $\chi$ is equivalent to the statistic of a noisy state $\chi^{\eta}$ and noiseless measurement, i.e.,

  \begin{equation} \label{equivlhv}
    \Tr( M_a^\eta \otimes N_b^\eta \chi) = \Tr(M_a \otimes N_b \chi^\eta),
  \end{equation}
  where $\chi^\eta$ is found by applying the dual map twice (once on Alice's side and once on Bob's side), namely,
  \begin{equation}\label{dualmap2}
    M_{*}^{\eta}(\chi)  = \eta^2 \chi + \eta (1-\eta) [ \chi_A \otimes \xi + \xi \otimes \chi_B] +(1-\eta)^2 \xi \otimes \xi
  \end{equation}
  where $\xi$ is the density matrix defining the map (see Eq.~\eqref{map}) and $\chi_A$, $\chi_B$ are the reduced state of $\chi$. Note that one could in principle extend the above equality to the case of different noisy maps for Alice and Bob.

  The final step is to prove that the left-hand side of Eq.~\eqref{equivlhv} admits a LHV model, which implies that the right-hand side also does. This can be done by considering only finitely many measurements, which form a polytope that contains the noisy set defined in Eq.~\eqref{map}, which makes it enough to focus on the extremal points. We formalize this idea by defining the ``shrinking factor'': given a set of measurements $\mathcal{M}$, a finite set $\{M_{a|x} \}$, and a map $\Phi^\eta$ we call the shrinking factor the largest value of $\eta$ such that $\mathcal{M}^\eta$ is included in the polytope defined by $\{M_{a|x} \}$. For a general method to compute the shrinking factor see \cite{Hirsch16}, Appendix A.

  To summarize the procedure: given a continuous set of measurements $\mathcal{M}$ and a noisy map $\Phi^{\eta}$ one finds a finite set, say $\{M_{a|x} \}$, which includes the noisy set $\mathcal{M}^\eta$ (for some fixed $\eta$). Then one finds $\chi$, such that its statistic with the finite sets is local and such that $\chi^\eta=\rho$, hence proving that $\rho$ is local for the continuous set $\mathcal{M}$. This procedure can be done in two distinct steps: first choose two noisy maps (\ie set the maps parameters $\xi_A$ and $\xi_B$), take finite sets $ \{M_{a|x} \}$ and $ \{N_{b|y} \}$ (with respective associated shrinking factors $\eta$ and $\mu$) and solve the following linear problem:
  \medbreak
  {\bf LHV Protocol}
  \ba \label{LHVprot} \text{find  } & & q^* = \max  q  \\
  \text{s.t.  }   & &\Tr(M _{a|x} \otimes N_{b|y} \chi) = \sum_\lambda p_\lambda D_\lambda(ab|xy)\quad \forall a,b,x,y   \nonumber  \\ 
  & & p_\lambda \geq 0 \quad \forall \lambda  \nonumber \\
  & &  q \rho + (1-q) \frac{\mathds{1}}{d}    =  \eta \mu \chi + \eta (1-\mu) \chi_A \otimes \xi_B  \nonumber \\ \nonumber
  & &  \quad \quad  \quad \quad  +\mu (1-\eta) \xi_A \otimes \chi_B +(1-\eta) (1-\mu) \xi_A \otimes \xi_B
  \ea
  where the optimization variable are (i) positive coefficients $p_\lambda$ and (ii) a $d_A \times d_B$ hermitian matrix $\chi$. Given $m_A$ ($m_B$) measurements with $k_A$ ($k_B$) outcomes for Alice (Bob), one has $n = (k_A)^{m_A} (k_B)^{m_B}$ local deterministic strategies $D_\lambda(ab|xy)$, and $\lambda = 1,\ldots,n$.

  If the result of the optimization $q^*$ satisfy $q^* > 1$ we have then ensured the existence of a LHV model for $\rho$, otherwise, we have ensured the existence of a LHV model for $  q \rho + (1-q) \mathds{1}/d$, with $q \leq q^*$, and we can repeat the procedure using larger sets $ \{M'_{a|x} \}$ and $ \{N'_{b|y} \}$, until $q^* \geq 1$. One can prove that this will happen if there exists a LHV model for $\rho$, \ie all local states are eventually detected by this algorithmic method. The precise iterative procedure is given below.

  Start from a finite set of measurements $\{ M^0_{a|x} \}$. Compute the shrinking factor $\eta_0$ of this finite set with respect to all POVMs of dimension $d_A$ (and with the desired number of outcomes),\footnote{We need $\eta_0 > 0$ for the sequence to converge, which can always be done by choosing a simplex as a first set} and for some $\xi_A$. This is the initial settings ($k=0$) of the following iterative process:

  \medbreak

  Step 1: Take measurements $\{ M^k_{a|x} \}$ and run the LHS protocol \eqref{LHSprot} on the family of states $\rho_p = p \rho + (1-p) \mathds{1}_D / D$, where $D = d_A \cdot d_B$.
  \begin{itemize} \label{iterative2}
    \item If $q^* \geq 1$ the algorithm stops and returns $R=0$ together with the values of the SDP variables. This ensures that $\rho$ admits a LHS model.
    \item If $q^* < 1$, we construct another finite set of measurements $\{ M^{k+1}_{a|x} \}$ with associated shrinking factor $\eta_{k+1} > \eta_k$. A way to do it is simply by taking $\{ M^k_{a|x} \}$ and adding the measurements which maximize the scalar product with the facets of $\{ M^k_{a|x} \}$, that is, the measurements which are ``furthest'' away from the set $\{ M^k_{a|x} \}$.\footnote{We can find the POVM violating maximally a facet thanks to the fact that the set has a SDP-characterization. This works for the set of all POVMs of a given dimension and number of outcomes, but fails for projective measurements, where the condition $P^2=P$ is not a SDP condition.}
  \end{itemize}
  Step 2: Set $k=k+1$ and go back to step $1$.

  One can also apply the improvements discussed in Section \ref{opt} which leads to the following protocol:
  \medbreak
  {\bf LHV Protocol.} (final version)
  \ba \label{LHVprotim} \text{find  } & &  q^* = \max  q  \\
  \text{s.t. }   & & \Tr(M _{a|x} \otimes N_{b|y} \chi) = \sum_\lambda p_\lambda D_\lambda(ab|xy) \quad \forall a,b,x,y \nonumber \\
  & & p_\lambda \geq 0\quad \forall \lambda \nonumber \\
  & & \chi^{\nu,\mu} = \nu \mu \chi + \nu (1-\mu) \chi_A \otimes \xi_B+ \mu (1-\nu) \xi_A \otimes \chi_B \nonumber\\
  & & \qquad\quad + (1-\nu) (1-\mu)\Tr(\chi) \xi_A \otimes \xi_B  \nonumber \\
  & & \rho_q - \chi^{\nu,\mu} - \sum_k \beta_k \rho_k \geq 0 \nonumber \\
  & & (\rho_q  - \chi^{\nu,\mu} - \sum_k \beta_k \rho_k)^{T_B} \geq 0 \nonumber \\
  & & \Tr(\chi) \geq 0,\qquad \beta_k  \geq 0 \quad\forall k \nonumber
  \ea
  where the exponent ${T_B}$ stands for the partial transposition on Bob's side and the optimization variable are (i) positive coefficients $p_\lambda$ and $\beta_k$ and (ii) a $d \times d$ hermitian matrix $\chi$. Given $m_A$ ($m_B$) $o_A$-outcome ($o_B$-outcome) measurements, one has $n = (o^A)^{m_A} (o_B)^{m_B}$ local deterministic strategies $D_\lambda(ab|xy)$, and $\lambda = 1,\ldots,n$.

\end{appendix}


\begin{thebibliography}{10}

  \bibitem{Bell64} J.S.~Bell, {\it On the Einstein-Podolsky-Rosen paradox}, Physics {\bf 1}, 195--200 (1964).

  \bibitem{review} N.~Brunner, D.~Cavalcanti, S.~Pironio, V.~Scarani, and S.~Wehner, {\it Bell nonlocality}, Rev.~Mod.~Phys.~{\bf 86}, 419 (2014).

  \bibitem{ABG07} A.~Ac\'in, N.~Brunner, N.~Gisin, S.~Massar, S.~Pironio, and V.~Scarani, {\it Device-independent security of quantum cryptography against collective attacks}, Phys.~Rev.~Lett.~{\bf 98}, 230501 (2007).

  \bibitem{Col} R.~Colbeck, {\it Quantum and relativistic protocols for secure multi-party computation}, PhD thesis, University of Cambridge, ArXiv:0911.3814, (2008).

  \bibitem{PAM10} S.~Pironio, A.~Ac\'in, S.~Massar, A.~Boyer De La Giroday, N.D.~Matsukevich, P.~Maunz, S.~Olmschenk, D.~Hayes, L.~Luo, T.A.~Manning, and C.~Monroe, {\it Random numbers certified by Bell's theorem}, Nature {\bf 464}, 10 (2010).

  \bibitem{Rotem} R.~Arnon-Friedman, F.~Dupuis, O.~Fawzi, R.~Renner, and T.~Vidick, {\it Practical device-independent quantum cryptography via entropy accumulation}, Nat.~Commun.~{\bf 9}, 459 (2018).

  \bibitem{Werner89} R.F.~Werner, {\it Quantum states with Einstein-Podolsky-Rosen correlations admitting a hidden-variable model}, Phys.~Rev.~A {\bf 40}, 4277 (1989).

  \bibitem{Barrett02} J.~Barrett, {\it Nonsequential positive-operator-valued measurements on entangled mixed states do not always violate a Bell inequality}, Phys.~Rev.~A {\bf 65}, 042302 (2002).

  \bibitem{Acin06} A.~Ac\'{\i}n, N.~Gisin, and B.~Toner, {\it Grothendieck's constant and local models for noisy entangled quantum states}, Phys.~Rev.~A {\bf 73}, 062105 (2006).

  \bibitem{Almeida07} M.L.~Almeida, S.~Pironio, J.~Barrett, G.~T{\'o}th, and A.~Ac\'{\i}n, {\it Noise Robustness of the nonlocality of Entangled Quantum States}, Phys.~Rev.~Lett.~{\bf 99}, 040403 (2007).

  \bibitem{GHNL} F.~Hirsch, M.T.~Quintino, J.~Bowles, and N.~Brunner, {\it Genuine hidden quantum nonlocality}, Phys.~Rev.~Lett.~{\bf 111}, 160402, (2013).

  \bibitem{Augusiak_review} R.~Augusiak, M.~Demianowicz, and A.~Ac\'{\i}n, {\it Local hidden variable models for entangled quantum states},  J.~Phys.~A {\bf 42}, 424002 (2014).

  \bibitem{Toth06} G.~Toth and A.~Ac\'{\i}n, {\it Genuine tripartite entangled states with a local hidden-variable model}, Phys.~Rev.~A {\bf 74}, 030306 (2006).

  \bibitem{Augusiak15} R.~Augusiak, M.~Demianowicz, J.~Tura, and A.~Ac\'{\i}n, {\it Entanglement and nonlocality are inequivalent for any number of parties}, Phys.~Rev.~Lett.~{\bf 115}, 030404 (2015).

  \bibitem{Bowles16b} J.~Bowles, J.~Francfort, M.~Fillettaz, F.~Hirsch, and N.~Brunner, {\it Genuinely multipartite entangled quantum states with fully local hidden variable models and hidden multipartite nonlocality}, Phys.~Rev.~Lett.~{\bf 116}, 130401 (2016).

  \bibitem{Wiseman07} H.M.~Wiseman, S.J.~Jones, and A.C.~Doherty, {\it Steering, entanglement, nonlocality, and the Einstein-Podolsky-Rosen paradox}, Phys.~Rev.~Lett.~{\bf 98}, 140402 (2007).

  \bibitem{paul_review} D.~Cavalcanti, P.~Skrzypczyk, {\it Quantum steering: a review with focus on semidefinite programming}, Rep.~Prog.~Phys.~{\bf 80}, 024001 (2017).

  \bibitem{Quintino15} M.T.~Quintino, T.~V\'ertesi, D.~Cavalcanti, R.~Augusiak, M.~Demianowicz, A.~Ac\'{\i}n, and N.~Brunner, {\it Inequivalence of entanglement, steering, and Bell nonlocality for general measurements}, Phys.~Rev.~A {\bf 92}, 032107 (2015).

  \bibitem{Bowles14} J.~Bowles, T.~V\'ertesi, M.T.~Quintino, and N.~Brunner, {\it One-way Einstein-Podolsky-Rosen steering}, Phys.~Rev.~Lett.~{\bf 112}, 200402 (2014).

  \bibitem{Jevtic15} S.~Jevtic, M.J.~W.~Hall, M.R.~Anderson, M.~Zwierz, and H.M.~Wiseman, {\it Einstein-Podolsky-Rosen steering and the steering ellipsoid}, Journal of the Optical Society of America B Optical Physics {\bf 32}, A40 (2015).

  \bibitem{Bowles15} J.~Bowles, F.~Hirsch, M.T.~Quintino, and N.~Brunner, {\it Local hidden variable models for entangled quantum states using finite shared randomness}, Phys.~Rev.~Lett.~{\bf 114}, 120401 (2015).

  \bibitem{Bowles16} J.~Bowles, F.~Hirsch, M.T.~Quintino, and N.~Brunner, {\it Sufficient criterion for guaranteeing that a two-qubit state is unsteerable}, Phys.~Rev.~A {\bf 93}, 022121 (2016).

  \bibitem{Nguyen16} H.~Chau Nguyen and T.~Vu, {\it Necessary and sufficient condition for steerability of two-qubit states by the geometry of steering outcomes}, Europhysics Letters {\bf 115}, 10003 (2016),

  \bibitem{Miller18} C.A.~Miller, R.~Colbeck, and Y.~Shi, {\it Keyring models: an approach to steerability}, J.~Math.~Phys.~{\bf 59}, 022103 (2018).

  \bibitem{Uola14} R.~Uola, T.~Moroder, and O.~G{\"u}hne, {\it Joint measurability of generalized measurements implies classicality}, Phys.~Rev.~Lett.~{\bf 113}, 160403 (2014).

  \bibitem{Quintino14} M.T.~Quintino, T.~V{\'e}rtesi, and N.~Brunner, {\it Joint measurability, Einstein-Podolsky-Rosen steering, and Bell nonlocality}, Phys.~Rev.~Lett.~{\bf 113}, 160402 (2014).

  \bibitem{Uola15} R.~Uola, C.~Budroni, O.~G{\"u}hne, and J.P.~Pellonp{\"a}{\"a}, {\it A one-to-one mapping between steering and joint measurability problems}, Phys.~Rev.~Lett.~{\bfseries 115}, 230402 (2015).

  \bibitem{Hirsch16} F.~Hirsch, M.T.~Quintino, T.~V\'ertesi, M.F.~Pusey, and N.~Brunner, {\it Algorithmic construction of local hidden variable models for entangled quantum states}, Phys.~Rev.~Lett.~{\bf 117}, 190402 (2016).

  \bibitem{Cavalcanti16} D.~Cavalcanti, L.~Guerini, R.~Rabelo, and P.~Skrzypczyk, {\it General method for constructing local-hidden-variable models for entangled quantum states}, Phys.~Rev.~Lett.~{\bf 117}, 190401 (2016).

  \bibitem{Sainz15} A.B.~Sainz, N.~Brunner, D.~Cavalcanti, P.~Skrzypczyk, and T.~V\'ertesi, {\it Post-quantum steering}, Phys.~Rev.~Lett.~{\bf 115}, 190403 (2015).

  \bibitem{Hirsch16b} F.~Hirsch, M.T.~Quintino, J.~Bowles, T.~V\'ertesi, and N.~Brunner, {\it Entanglement without hidden nonlocality}, New J.~Phys.~{\bf 18}, 113019 (2016).

  \bibitem{Nagy16} S.~Nagy and T.~V\'ertesi, {\it EPR Steering inequalities with Communication Assistance}, Sci.~Rep.~{\bf 6}, 21634 (2016).

  \bibitem{Bavaresco17} J.~Bavaresco, M.T.~Quintino, L.~Guerini, T.O.~Maciel, D.~Cavalcanti, and M.~Terra Cunha, {\it Most incompatible measurements for robust steering tests}, Phys.~Rev.~A {\bf 96}, 022110 (2017).

  \bibitem{Hirsch17} F.~Hirsch, M.T.~Quintino, T.~V{\'e}rtesi, M.~Navascu\'es, and N.~Brunner, {\it Better local hidden variable models for two-qubit Werner states and an upper bound on the Grothendieck constant KG(3)}, Quantum {\bf 1}, 3 (2017).

  \bibitem{Toth18} G.~Toth and T.~V\'ertesi, {\it Quantum states with a positive partial transpose are useful for metrology}, Phys.~Rev.~Lett.~{\bf 120}, 020506 (2018).

  \bibitem{Orieux17} A.~Orieux, M.~Kaplan, V.~Venuti, T.~Pramanik, I.~Zaquine, and E.~Diamanti, {\it Experimental detection of steerability in Bell local states with two measurement settings}, J.~Opt.~{\bf 20}, 044006 (2018).

  \bibitem{Bene18} E.~Bene and T.~V\'ertesi, {\it Measurement incompatibility does not give rise to Bell violation in general}, New J.~Phys.~{\bf 20}, 013021 (2018).

  \bibitem{Hirsch18} F.~Hirsch, M.T.~Quintino, and N.~Brunner, {\it Quantum measurement incompatibility does not imply Bell nonlocality}, Phys.~Rev.~A {\bf 97}, 012129 (2018).

  \bibitem{Zhang17} F.L.~Zhang and Y.Y.~Zhang, {\it Local hidden state models for Bell diagonal states}, arXiv:1709.09124.

  \bibitem{Yu17} B.C.~Yu, Z.A.~Jia, Y.~Wu, and G.C.~Guo, {\it Geometric local hidden state model for some two-qubit states}, arXiv:1710.06704.

  \bibitem{Terhal} B.M.~Terhal, A.C.~Doherty, and D.~Schwab, {\it Symmetric extensions of quantum states and local hidden variable theories}, Phys.~Rev.~Lett.~{\bf 90}, 157903 (2003).

  \bibitem{Pusey13} M.F.~Pusey, {\it Negativity and steering: a stronger Peres conjecture}, Phys.~Rev.~A {\bf 88}, 032313 (2013).

  \bibitem{Paul14} P.~Skrzypczyk, M.~Navascu\'es, and D.~Cavalcanti, {\it Quantifying Einstein-Podolsky-Rosen steering}, Phys.~Rev.~Lett.~{\bf 112}, 180404 (2014).

  \bibitem{Dariano05} G.~Mauro D'Ariano, P.~Lo Presti, and P.~Perinotti, {\it Classical randomness in quantum measurements}, J.~Phys.~A: Math.~Gen.~{\bf 38}, 5979--5991 (2005).

  \bibitem{Peres96} A.~Peres, {\it Separability Criterion for Density Matrices}, Phys.~Rev.~Lett.~{\bf 77}, 1413 (1996).

  \bibitem{Horodecki96} M.~Horodecki, P.~Horodecki, and R.~Horodecki, {\it Separability of mixed states: necessary and sufficient conditions}, Phys.~Lett.~A {\bf 223}, 1 (1996).

  \bibitem{Horodecki97} P.~Horodecki, {\it Separability criterion and inseparable mixed states with positive partial transposition}, Phys.~Lett.~A {\bf 232}, 333 (1997).

  \bibitem{Moroder14} T.~Moroder, O.~Gittsovich, M.~Huber, and O.~Guhne, Phys.~Rev.~Lett.~{\bf 113}, 050404 (2014).

  \bibitem{Vertesi14} T.~V\'ertesi and N.~Brunner, {\it Disproving the Peres conjecture: Bell nonlocality from bipartite bound entanglement}, Nat.~Commun.~{\bf 5}, 5297 (2014).

  \bibitem{Rosset14} D.~Rosset, J.D.~Bancal, and N.~Gisin, {\it Classifying 50 years of Bell inequalities}, J.~Phys.~A: Math.~Theor.~{\bf 47}, 424022 (2014).

  \bibitem{Renou17} M.O.~Renou, D.~Rosset, A.~Martin, and N.~Gisin, {\it On the inequivalence of the CH and CHSH inequalities due to finite statistics}, J.~Phys.~A: Math.~Theor.~{\bf 50}, 255301 (2017).

  \bibitem{Nguyen17} H.C.~Nguyen, A.~Milne, T.~Vu, and S.~Jevtic, {\it Quantum steering with positive operator valued measures}, arXiv:1706.08166.

  \bibitem{Acin02} A.~Acin, T.~Durt, N.~Gisin, and J.I.~Latorre, {\it Quantum non-locality in two three-level systems}, Phys.~Rev.~A {\bf 65}, 052325 (2002).

  \bibitem{Toner03} B.F.~Toner and D.~Bacon, {\it Communication cost of simulating Bell correlations}, Phys.~Rev.~Lett.~{\bf 91}, 187904 (2003).

  \bibitem{PR} S.~Popescu and D.~Rohrlich, {\it Quantum nonlocality as an axiom}, Found.~Phys.~{\bf 24}, 379 (1994).

  \bibitem{Cerf05} N.J.~Cerf, N.~Gisin, S.~Massar, and S.~Popescu, {\it Simulating maximal quantum entanglement without communication}, Phys.~Rev.~Lett.~{\bf 94}, 220403 (2005).

  \bibitem{Brunner05} N.~Brunner, N.~Gisin, and V.~Scarani, {\it Entanglement and nonlocality are different resources}, New J.~Phys.~{\bf 7}, 88 (2005).

\end{thebibliography}
\end{document}